\documentclass[useAMS,usenatbib,referee]{biom}
\usepackage{amsmath}
\usepackage{graphicx,psfrag,epsf}
\usepackage{enumerate}
\usepackage{natbib}
\usepackage{hyperref}
\usepackage{float}
\usepackage{subcaption}

\DeclareMathOperator*{\argmin}{argmin}
\usepackage{enumitem}
\usepackage{mathrsfs}
\usepackage{amsfonts}
\usepackage{multirow}
\usepackage{rotating}
\usepackage{xr}
\makeatletter
\newcommand*{\addFileDependency}[1]{
  \typeout{(#1)}
  \@addtofilelist{#1}
  \IfFileExists{#1}{}{\typeout{No file #1.}}
}
\makeatother

\newcommand*{\myexternaldocument}[1]{
    \externaldocument{#1}
    \addFileDependency{#1.tex}
    \addFileDependency{#1.aux}
}
\myexternaldocument{FRASLD_supplement_v2}
\def\bSig\mathbf{\Sigma}

\newcommand{\tr}{\mbox{tr}}

\def\wt{\widetilde}
\renewcommand{\theequation}{\thesection.\arabic{equation}}
\renewcommand{\theequation}{\arabic{equation}}
\renewcommand{\theequation}{\arabic{equation}}
\def\argmin{\mathrm{argmin\ }}

\def\bse{\begin{eqnarray*}}
	\def\ese{\end{eqnarray*}}
\def\be{\begin{eqnarray}}
\def\ee{\end{eqnarray}}
\newcommand{\ben}{\begin{eqnarray}}\newcommand{\een}{\end{eqnarray}}
\newcommand{\bn}{\begin{enumerate}}
	\newcommand{\en}{\end{enumerate}}

\newcommand{\bc}{\begin{cases}}
	\newcommand{\ec}{\end{cases}}
\newcommand{\bt}{\begin{tabular}}
	\newcommand{\et}{\end{tabular}}
\newcommand{\bct}{\begin{center}}
	\newcommand{\ect}{\end{center}}
\def\wt{\widetilde}
\def\diag{\hbox{diag}}
\def\wh{\widehat}

\newcommand{\cid}{\buildrel d \over \longrightarrow}

\newcommand{\bdb}{{\bf b}}

\newcommand{\vertiii}[1]{{\left\vert\kern-0.25ex\left\vert\kern-0.25ex\left\vert #1 
		\right\vert\kern-0.25ex\right\vert\kern-0.25ex\right\vert}}

\def\trans{^{\rm T}}

\newcommand{\bone}{{\bf 1}}

\newcommand{\bdepsilon}{{\pmb \epsilon}}

\newcommand{\bdmu}{{\pmb \mu}}
\newcommand{\bdpsi}{{\pmb \psi}}

\newcommand{\bdbeta}{{\pmb \beta}}

\newcommand{\bdxi}{{\pmb \xi}}

\newcommand{\CA}{{\cal A}}

\newcommand{\CQ}{{\cal Q}}

\newcommand{\CT}{{\cal T}}

\newcommand{\BI}{{\bf I}}

\newcommand{\BS}{{\bf S}}
\newcommand{\BT}{{\bf T}}
\newcommand{\BU}{{\bf U}}

\newcommand{\BW}{{\bf W}}

\newcommand{\BX}{{ \bf X }} % on t domain
\newcommand{\BY}{{ \bf Y}} % on t domain
\newcommand{\BZ}{{\bf Z}}

\newcommand{\mti}{{m_{y,i}}}
\newcommand{\msi}{{m_{x,i}}}

\newcommand{\MX}{{\mathbb X}}

\newcommand{\BSigma}{{\bf \Sigma}}
\newcommand{\BLambda}{{\bf \Lambda}}
\newcommand{\BPsi}{{\bf \Psi}} % on s domain
\newcommand{\BPsit}{{{\bf \Psi}_*}} % on t domain

\newcommand{\BOmega}{{\pmb \Omega}}

\newcommand{\Normal}{\hbox{Normal}}

\newcommand{\E}{{\rm E}}
\newcommand{\cov}{{\rm Cov}}
\newcommand{\var}{{\rm Var}}

\def\II{I\negthinspace I}

\usepackage{xcolor}
\newcommand{\xc}[1]{{\color{black}#1}}
\newcommand{\yhl}[1]{\textcolor{black}{#1}}
\newcommand{\yl}[1]{\textcolor{black}{#1}}

\allowdisplaybreaks

\title[Functional Calibration]{ Asynchronous and Error-prone Longitudinal Data Analysis via Functional Calibration}

\author{Xinyue Chang$^{1,*}$\email{xchang225@gmail.com}, 
Yehua Li$^{2,**}$\email{yehuali@ucr.edu}, and 
Yi Li$^{3,***}$\email{yili@umich.edu} \\
$^{1}$Department of Statistics, Iowa State University, Ames, IA 50011, U.S.A. \\
$^{2}$Department of Statistics, University of California, Riverside, CA 92521, U.S.A. \\
$^{3}$Department of Biostatistics, University of Michigan, Ann Arbor, MI 48109, U.S.A.}

\begin{document}

%  This will produce the submission and review information that appears
%  right after the reference section.  Of course, it will be unknown when
%  you submit your paper, so you can either leave this out or put in 
%  sample dates (these will have no effect on the fate of your paper in the
%  review process!)

\date{{\it Received October} 2007. {\it Revised February} 2008.  {\it
Accepted March} 2008.}

%  These options will count the number of pages and provide volume
%  and date information in the upper left hand corner of the top of the 
%  first page as in published papers.  The \pagerange command will only
%  work if you place the command \label{firstpage} near the beginning
%  of the document and \label{lastpage} at the end of the document, as we
%  have done in this template.

%  Again, putting a volume number and date is for your own amusement and
%  has no bearing on what actually happens to your paper!  

\pagerange{\pageref{firstpage}--\pageref{lastpage}} 
\volume{64}
\pubyear{2022}
\artmonth{October}

%  The \doi command is where the DOI for your paper would be placed should it
%  be published.  Again, if you make one up and stick it here, it means 
%  nothing!

\doi{10.1111/j.1541-0420.2005.00454.x}

%  This label and the label ``lastpage'' are used by the \pagerange
%  command above to give the page range for the article.  You may have 
%  to process the document twice to get this to match up with what you 
%  expect.  When using the referee option, this will not count the pages
%  with tables and figures.  

\label{firstpage}

%  put the summary for your paper here

\begin{abstract}
In many longitudinal settings,  time-varying covariates may not be measured at the same time as responses and are often prone to measurement error. Naive last-observation-carried-forward methods incur estimation biases, and existing kernel-based methods suffer from slow convergence rates and large variations. 
{To address these challenges, we propose a new functional calibration approach to efficiently learn  longitudinal covariate processes based on sparse functional data with measurement error. Our approach, stemming from functional principal component analysis, calibrates the unobserved synchronized covariate values from the observed asynchronous and error-prone covariate values, 
and is broadly applicable} to asynchronous longitudinal regression with  time-invariant or time-varying coefficients. For regression with time-invariant coefficients, our estimator is asymptotically unbiased, root-n consistent, and asymptotically normal; for time-varying coefficient models, our estimator has the optimal varying coefficient model convergence rate with inflated asymptotic variance from the calibration. In both cases, our estimators present asymptotic properties superior to the existing methods. The feasibility and usability of the proposed methods are verified by simulations and an application to the Study of Women's Health Across the Nation, a large-scale multi-site longitudinal study on women's health during  mid-life.
\end{abstract}

%  Please place your key words in alphabetical order, separated
%  by semicolons, with the first letter of the first word capitalized,
%  and a period at the end of the list.
%


%  As usual, the \maketitle command creates the title and author/affiliations
%  display 

\maketitle

%  If you are using the referee option, a new page, numbered page 1, will
%  start after the summary and keywords.  The page numbers thus count the
%  number of pages of your manuscript in the preferred submission style.
%  Remember, ``Normally, regular papers exceeding 25 pages and Reader Reaction 
%  papers exceeding 12 pages in (the preferred style) will be returned to 
%  the authors without review. The page limit includes acknowledgements, 
%  references, and appendices, but not tables and figures. The page count does 
%  not include the title page and abstract. A maximum of six (6) tables or 
%  figures combined is often required.''

%  You may now place the substance of your manuscript here.  Please use
%  the \section, \subsection, etc commands as described in the user guide.
%  Please use \label and \ref commands to cross-reference sections, equations,
%  tables, figures, etc.
%
%  Please DO NOT attempt to reformat the style of equation numbering!
%  For that matter, please do not attempt to redefine anything!

\section{Introduction}
\label{sec:introduction}
\begin{figure}[ht]
	\centering
	\begin{tabular}{c}	
    \includegraphics[width = 5in, height=3in]{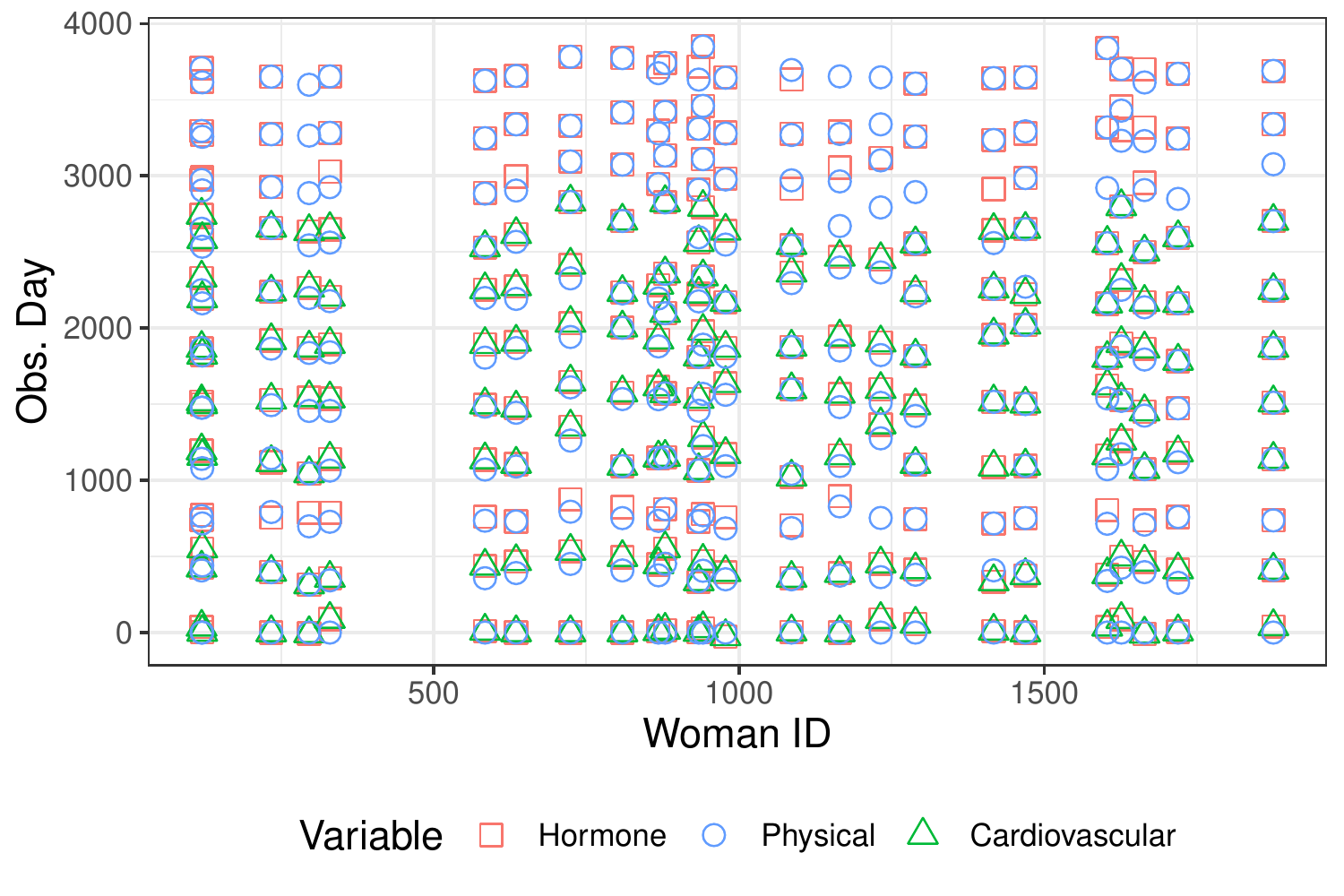} \\
    (a)\\
    \includegraphics[width = 5in, height=2.5in]{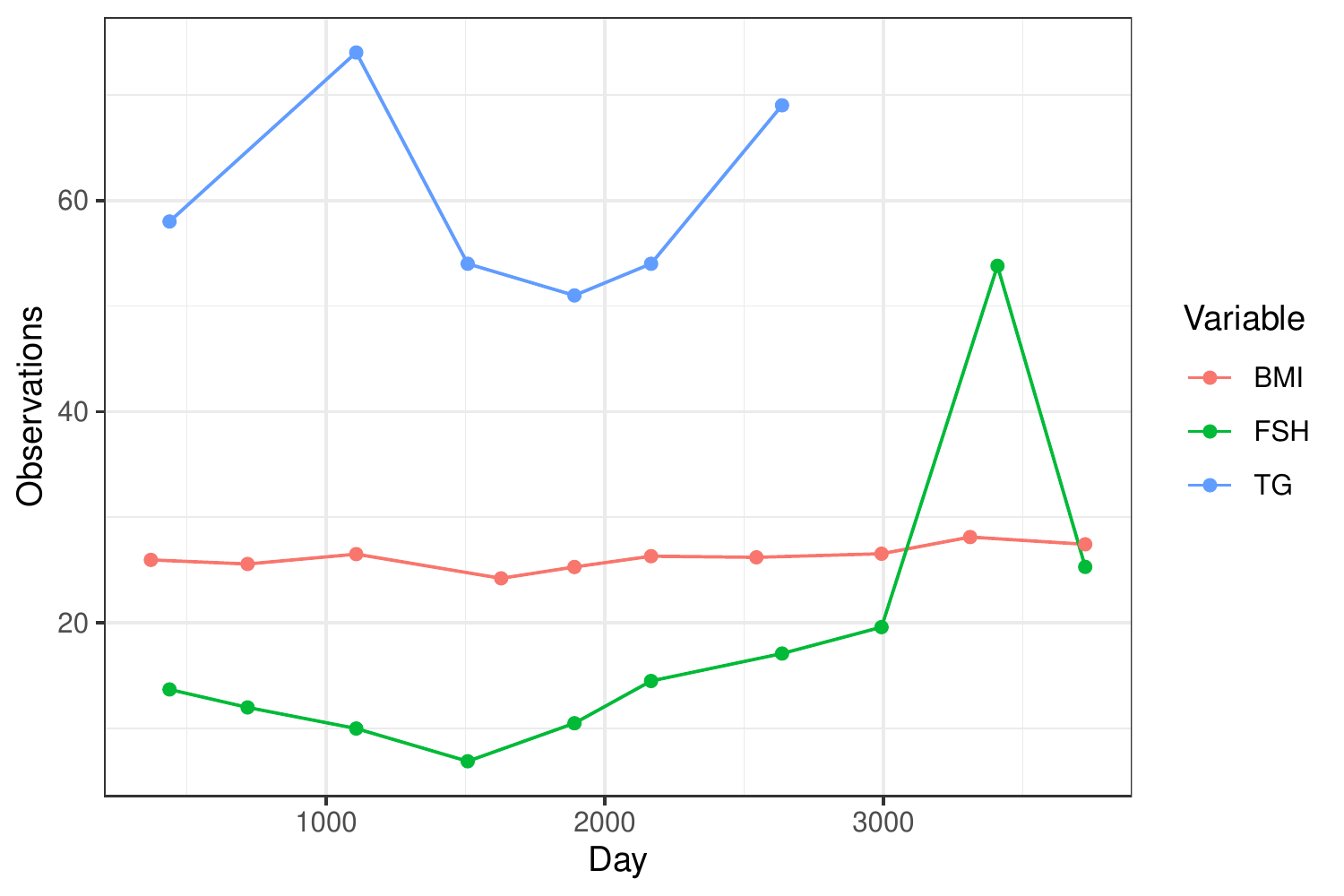}\\
    (b)
    \end{tabular}
    \vspace{1cm}
	\caption[Observation days for different women participants in SWAN.]{(a) Observation days for a randomly selected subset of SWAN participants. Each column corresponds to one woman, with points in different colors and shapes representing variables types: hormone, physical and cardiovascular measures. (b) Longitudinal trajectories of follicle-stimulating hormone (FSH), body mass index (BMI), and triglycerides (TG) from a randomly selected SWAN participant (ID = 13959).}
	\label{fig:days_id}
\end{figure}

%In many large longitudinal studies, variables of interest are collected in different databases and not synchronized in time. 
In many decade-long longitudinal studies,  participants' heath information  
is repeatedly measured by diverse instruments, such as blood tests, physical examinations, nutritional evaluations and psychological assessments. These tests and assessments usually follow  different schedules, and are not synchronized in time. The resulting data structures create an asynchronous issue where the response variable and covariates are not measured at the same time.
{For example, in our motivating Study of Women's Health Across the Nation (SWAN; \url{https://www.swanstudy.org/}), a  multi-site longitudinal study on women's health during their mid-life years,  a total of 3,302 women were followed from 1996 to 2008 to study their physical, biological, psychological, and social changes that occurred during the menopausal transition.} These health-related metrics were grouped into physical, hormone, and cardiovascular measurements.  Fig \ref{fig:days_id}(a) shows the measurement times for hormone, physical and cardiovascular measurements for a random sample of SWAN participants. As  seen, these measurements were taken following  different schedules. During this important transition, of particular interest is the level of the follicle-stimulating hormone (FSH), our response variable. Two important physical and cardiovascular covariates, the body mass index (BMI) and triglycerides (TG), are also repeatedly measured but on different schedules. {Another complication as manifested by Fig \ref{fig:days_id}(b),  a spaghetti plot for the longitudinal trajectories of these three variables from a randomly selected participant, is that these asynchronized variables also exhibit short term fluctuations, which need to be modeled as measurement error or nugget effect \citep{Carroll2006}.}

%A naive last-observation-carried-forward (LOCF) method ignores the dynamics of time-varying covariates, takes the most recent measurement as the current value of the covariate, and performs routine regression analyses. It is known such a method leads to biased and inconsistent estimates of the regression coefficients \citep{Molenberghs2004biostatistics}. 
There has been some literature on analyzing incomplete longitudinal data using missing data techniques such as the inverse probability weighting: \cite{Robins1995} assumed the response and time-varying covariates must be missing or present simultaneously; \cite{Cook2004} assumed that the repeated measurements within a subject are complete before the subject dropout from the study. These methods rely on parametric modeling of the missing data mechanism and are not designed for data that are asynchronous by design.

More recently, \cite{cao2015regression} 
%and \cite{Cao2016ejs} 
modeled asynchronous longitudinal data under generalized linear models with either time-invariant or time-varying coefficients, by proposing kernel-weighted estimating equation methods to down-weight covariates that are further away in time from the response. These kernel-weighted estimators are consistent and asymptotically normal, but with slow convergence rates. For time-invariant regression models, their estimated regression coefficients converge in a nonparametric kernel regression rate instead of the usual root-$n$ parametric rate; for time-varying coefficient models, their estimator converges in a bivariate nonparametric smoothing rate, which is much slower than classic convergence rate of varying coefficient models in \cite{Cai2000jasa} and sensitive to bandwidth selection, as shown in our simulation studies. {Also, %to our knowledge, 
none of the existing methods
adequately address the measurement error issue arising from
the asynchronous variables.
}

To address these limitations, we propose to model the longitudinal trajectories of the covariates in the SWAN study as functional data \citep{Ramsay2005}, and use the functional principal component analysis (FPCA) technique 
%\citep{yao2005functional, 
\citep{li2010uniform}
%, zhang2016sparse} 
to impute the missing synchronized covariate values from the observed asynchronous, error-prone covariate values. We then use the imputed values in second stage regression analyses. This method is similar in spirit to the regression calibration method in the measurement error literature \citep{Carroll2006}, but is completely nonparametric. We therefore term our proposed methodology Functional Calibration for Asynchronous Regression (FCAR). %  "as" is not needed following term 
The proposed method can be easily implemented using existing software, such as the `fdapace' package in R, and combined with other existing regression methodology such as the common linear regression with time-invariant regression coefficients and the time-varying coefficients \citep{hoover1998nonparametric}. We show that our estimators for time-invariant coefficient regression models are root-$n$ consistent and asymptotic normal, while our estimators for time-varying coefficient models enjoy the optimal convergent rate as \cite{hoover1998nonparametric}, 
%Cai2000jasa}, 
which is one order of magnitude faster than the existing methods such as \cite{cao2015regression}. We also show that our method can be  extended to accommodate multiple asynchronous longitudinal covariate processes using multivariate FPCA \citep{%chiou2014multivariate,  
happ2018multivariate, dai2021modeling}.

We are aware of the related work on asynchronous longitudinal regression using functional data analysis approaches. For example, \cite{csenturk2010functional} and \cite{csenturk2013modeling} proposed to estimate the time-varying coefficient function by estimating the covariance function of the time-varying covariate and cross-covariance function between the covariate and response processes using bivariate kernel smoothing. However, their method is associated with the same slower bivariate smoothing convergence rate as \cite{cao2015regression}. Our simulation study shows our method outperforms \cite{csenturk2013modeling} and \cite{cao2015regression} in efficiency and numerical stability. 

The paper is organized as follows. We introduce our model assumptions in Section \ref{sec:notations}, propose the new functional calibration method and apply it to longitudinal regression models with time-invariant or time-varying coefficients in Section \ref{sec:methods}. The asymptotic properties of the proposed estimators are established in Section \ref{sec:asymptotics}, while the practical performance of the proposed methods is illustrated by simulation studies in Section \ref{sec:simulation}. We apply the proposed methods to analyze the SWAN data and investigate the potential effects of BMI and triglycerides on follicle-stimulating hormone changes during menopausal transition in Section \ref{sec:realdata}. We provide concluding remarks  in Section \ref{sec:conclusion}. We   present technical proofs and additional numerical results (tables and graphs) and extend the proposed method to multivariate time-varying covariate processes  in the  Appendix.

\section{Model Assumptions} \label{sec:notations}
Let $\{X_i(t), Y_i(t)\}$, $i=1,\ldots, n$, be independent and identically distributed (iid) bivariate longitudinal processes defined on a compact time interval $\mathcal{T} \subset \mathbb{R}$, where $Y_i(t)$ is the response of the $i$th subject at time $t$ and $X_i(t)$ is the time-varying covariate process. For simplicity, we focus on the case where $X_i(t)$ is a univariate process, and present its multivariate extension in Appendix B.
%Following \cite{cao2015regression},
We will consider both the time-invariant coefficient model  
\be \label{ti_model}
	Y_i(t)= \beta_0 + \beta_1X_i(t) + \epsilon_i(t), 
\ee
and the time-varying coefficient model 
\be \label{tv_model}
	Y_i(t) = \beta_0(t) + \beta_1(t)X_i(t) + \epsilon_i(t).
\ee
In Model (\ref{ti_model}), $\bdbeta=(\beta_0, \beta_1)\trans$ are the time-invariant intercept and slope parameters, while $\beta_0(\cdot)$ and $\beta_1(\cdot)$ are the time-varying counterparts in Model (\ref{tv_model}). In both models, we assume $\epsilon_i(t)$ are iid zero-mean error processes with covariance function $\Omega(s,t)= \cov\{ \epsilon (s), \epsilon(t)\}$. We also assume that $X_i(t)$ and $\epsilon_i(t)$ are independent of each other.

Longitudinal variables are observed on discrete time points. Denote by  $\BT_i = (T_{i1}, \ldots, T_{i\mti})\trans$  the time points when $Y_i(\cdot)$ is observed, and  by $\BY_i=(Y_{i1}, \ldots, Y_{i\mti})\trans$ the observed response vector, where $Y_{ij}=Y_i(T_{ij})$, $j=1,\ldots, \mti$.
On the other hand, in an asynchronous longitudinal design, $X(t)$ are observed on time points $\BS_i = (S_{i1}, \ldots, S_{i\msi})\trans$. Let $\BX_i=(X_{i1}, \ldots, X_{i \msi})\trans$ where $X_{ij}=X_i(S_{ij})$, $j=1, \ldots, \msi$. As illustrated in our motivating example,  $\BS_i$ and $\msi$ can be completely different from $\BT_i$ and $\mti$. In addition, these time-varying covariates are usually contaminated with measurement error \citep{liao2011survival}. These measurement errors,  not necessarily arising from instrument error, may be due to local variations. In the SWAN study, as the BMI and triglycerides level naturally {fluctuate over days or even within the same day, it is reasonable to use their long-term or average values as the ``true" values}
to predict the response; failing to take into account the measurement error can result in biased estimates and reduced statistical power \citep{Carroll2006}. To proceed, we relate $X_{ij}$, the ``truth", to its  observed error-contaminated surrogates, $W_{ij}$, via an additive measurement error model: %\citep{Carroll2006}:}
\be\label{eq:disc_obs}
	W_{ij} = X_{ij} + U_{ij}, \quad j = 1, \ldots, \msi, \quad i = 1, \ldots, n, 
\ee
where $U_{ij}$ are iid zero-mean measurement errors with variance $\sigma_u^2$ and independent of $\BX_i$.  Model (\ref{ti_model}) or   (\ref{tv_model}), coupled with (\ref{eq:disc_obs}), is referred to as {\em asynchronous longitudinal regression with measurement error.}

Let $\BW_i=(W_{i1}, \ldots, W_{i \msi})\trans$ be the {\em observed surrogate} values that are {\em asynchronous} with $\BY_i$, whereas we denote by $\BX_{*i}=\{X_{i} (T_{i1}), \ldots, X_{i }(T_{i\mti})\}\trans$ the {\em unobserved true} covariate values that are {\em synchronized} with $\BY_i$. In the measurement error literature \citep{Carroll2006}, a commonly used technique to impute $\BX_{*i}$ from observed surrogate $\BW_i$ is regression calibration, which %usually relies on parametric models 
{ignores longitudinal correlations, does not capture the dynamic changes in the time-varying covariates and may incur efficiency loss. We instead propose to calibrate the value of $\BX_{*i}$ using a more efficient} functional data analysis approach.
We assume that the time-varying covariate $X(t), t \in \mathcal{T}$ is a stochastic process defined on $\mathcal{T}$ with mean and covariance functions 
\bse
	\mu(t) = \E\{X_i(t)\}, \quad R(s, t) = \cov\{X_i(s), X_i(t)\}, \quad s, t, \in \mathcal{T}.
\ese
The covariance function is a smooth, symmetric, positive semi-definite function with a spectral decomposition of $R(s,t)=\sum_{k=1}^q \omega_k \psi_k(s) \psi_k(t)$, where $\omega_1\ge \omega_2\ge \cdots \ge \omega_q > 0$ are the eigenvalues, and $\psi_k(\cdot)$ are the corresponding eigenfunctions (or  principal components) such that $\int_\CT \psi_k(t) \psi_{k'}(t) dt=I(k=k')$.
By the Karhunen--Lo\`{e}ve theorem, % \citep{Hsing-Eubank15}, 
\be \label{Xi_expansion}
	X_i(t) = \mu(t) + \hbox{$\sum_{k=1}^{q}$} \xi_{ik} \psi_k(t), \quad t \in \mathcal{T}, 
\ee
for $i = 1, \ldots, n$, where $\xi_{ik}=\int_\CT \{X_i(t) -\mu(t)\} \psi_k(t) dt$ are the principal component scores with mean zero and $\cov(\xi_{ik}, \xi_{ik'})= \omega_k I(k=k')$. 
The number of principal components $q$ can be infinity in theory, but
it is common to assume that $X_i(t)$ has a reduced rank representation with a finite $q$  %\citep{Zhou2010}
\citep{li2010uniform}. This is suitable for longitudinal or sparse functional data, where the number of measurements on each trajectories is so small that
one cannot realistically estimate a large number of principal components.
%It is common to assume that $X_i(t)$ has a reduced rank representation where $q$ is finite \citep{Zhou2010} and 
In practice, $q$ is chosen in a data-driven fashion  \citep{yao2005functional, li2013selecting}, which is to be detailed in Section \ref{sec:methods:functional_calibration}.

%Consider a typical longitudinal context, $Y_{ij}$ is the observed response at time point $T_{ij} \in \mathcal{T}$ of the $i$th subject, where $i = 1, \ldots, n$ and $j = 1, \ldots, m_{y,i}$. Under the asynchronous data framework where $T_{ij}$ and $S_{ij}$ are not necessarily the same, we assume the pair of observation time points, $(T_{ij}, S_{ij})$, are iid copies of a joint random variable $(T, S)$. As introduced initially, we consider sparsely observed longitudinal data where a limited number of observations are available per subject. Thus, we assume the number of observations for the covariate and response, $(\msi, \mti)$, are iid copies of a random variable bounded by $M < \infty$ almost surely.
%\begin{enumerate}[label=(A.\arabic*), resume]
%\end{enumerate}

%To study and assess cognition function for elderly women, we want to use a summarized score based on four memory impairment tests as the response variable \citep{weuve2004physical}. Meanwhile, we wanted to consider hormone and cardiovascular as independent variables, such as FSH and cholesterol.
%
%
%
%
%
%
%
%
%
%
%
%
%
%
%
%
%
%
%
%
\section{Functional Calibration for Asynchronous Regression} \label{sec:methods}
%The main idea of the functional calibration is to estimate functions $X_i(t), t \in \mathcal{T}$, $i = 1, \ldots, n$ via functional principal component analysis (FPCA), then $\wh X_i(T_{ij})$ is obtained to match with the response $Y_{ij}$. Thus, regression models can be applied directly to the synchronized longitudinal data achieved by functional calibration.

\subsection{Calibration using Functional Principal Component Analysis} \label{sec:methods:functional_calibration}
%Let function $f(t) \in L^2(\mathcal{T})$, define Hilbert space $\mathcal{H} = L^2(\mathcal{T})$ with inner product
%\bse
%\langle f, g\rangle = \int_{\mathcal{T}} f(t) g(t) dt,
%\ese
%where $f, g \in L^2(\mathcal{T})$. Then the norm on $\mathcal{H}$ induced by $\langle \cdot, \cdot \rangle$ is denoted as $||\cdot||$, contrast to $||\cdot||_2$ used as $\ell_2$ norm. Let $\mathfrak{R}$ be the corresponding integral operator of $R$, and $(\omega_k, \psi_k)$ are eigenvalue and eigenfunction pairs of $\mathfrak{R}$, i.e., 
%\bse
%(\mathfrak{R} \psi_k) (s) = \int_{\mathcal{T}} R(s, t) \psi_k(t) dt = \omega_k \psi_k(s), \quad s \in \mathcal{T}.
%\ese

Let $\bdmu_i = \{\mu(S_{i1}), \ldots, \mu(S_{i\msi})\}\trans $ and $\bdpsi_{ik} = \{\psi_k(S_{i1}), \ldots, \psi_k(S_{i\msi})\}\trans$, $k=1, \ldots, q$, be the mean and eigenfunctions interpolated on the observed time points, and put $\BPsi_{i} = [\bdpsi_{i1}, \ldots, \bdpsi_{iq}]$. Under the reduced rank model (\ref{Xi_expansion}) with a finite rank $q$, the within-subject covariance matrix for $\BW_i$ is $\BSigma_{i}=\cov(\BW_i)= \BPsi_i  \BLambda \BPsi_{i}\trans +\sigma_u^2 I $, where $\BLambda = \diag (\omega_1, \ldots, \omega_q)$. 
{If $\bdmu_i$ and $\BLambda$ were known, a roadmap for calibrating
 the unobserved, synchronized covariates $\BX_{*i}$ would be as follows.
First, the best linear unbiased predictors (BLUP) for the FPC scores would be}
\be\label{eq:fpc_score_blup}
	\wt \bdxi_i=(\wt \xi_{i1}, \ldots, \wt \xi_{iq})\trans= \BLambda \BPsi_i\trans \BSigma_{i}^{-1} (\BW_i-\bdmu_i).
\ee
Second, one could predict the functional trajectory of $X_i(t)$ by 
\ben\label{eq:X_tilde}
	\wt X_i(t) = \mu(t)+\hbox{$\sum_{k=1}^q$} \wt\xi_{ik} \psi_k(t).
\een
Finally, interpolating these predicted trajectories on the observation times of $Y$, we could predict the unobserved, synchronized covariates $\BX_{*i}$ by 
\ben\label{eq:X_tilde_star}
	\wt \BX_{*i}= \bdmu_{*i} +\hbox{$\sum_{k=1}^q$} \wt \xi_{ik} \bdpsi_{*ik} = \bdmu_{*i} +\BPsi_{*i} \wt \bdxi_i,
\een
where $\bdmu_{*i}=\{\mu(T_{i1}), \ldots, \mu(T_{i\mti})\}\trans$, $\BPsi_{*i} = \{\bdpsi_{*i1}, \ldots, \bdpsi_{*iq}\}$ and $\bdpsi_{*ik}= \{\psi_{k}(T_{i1}), \ldots,$ $\psi_{k}(T_{i\mti})\}\trans$, $k=1, \ldots, q$. 

{However, as $\bdmu_i$ and $\BLambda$ are  unknown,} and in order to complete this calibration roadmap, we  need to estimate {them}
%the mean and covariance function of $X(t)$ from 
based on the observed data $\{\BW_i, i=1, \ldots, n \}$. Let $K(\cdot)$ be a kernel function, and denote by $K_h(u) = K(u/h)/h $ where $h$ is the bandwidth. Following \cite{yao2005functional} and \cite{li2010uniform}, we use local linear smoothers to estimate mean and covariance functions. For any fixed $t$, we estimate $\mu(t)$ by $\wh \mu(t) = \wh a_0$, where
\[
	(\wh a_0, \wh a_1) = \underset{a_0, a_1}{\argmin} \dfrac{1}{n} \sum_{i=1}^n \dfrac{1}{\msi} \sum_{j=1}^{\msi} \{W_{ij} - a_0 - a_1 (S_{ij} - t)\}^2 K_{h_{\mu}} (S_{ij} - t), 
\]
with $h_\mu >0$ being the bandwidth. We then estimate $R(s,t)$ by $\wh R(s, t) = \wh a_0$ with
\bse
	(\wh a_0, \wh a_1, \wh a_2) &=& \underset{a_0, a_1, a_2}{\argmin} \dfrac{1}{n} \sum_{i=1}^n \bigg[\dfrac{1}{M_{x,i}} \sum_{j\neq l} \{L_{ij} L_{il} -a_0 - a_1(S_{ij} - s) - a_2 (S_{il} - t)\}^2 \\
	&& \hskip30mm \times K_{h_R} (S_{ij} - s) K_{h_R}(S_{il} - t) \bigg],
\ese
where $L_{ij} = W_{ij} - \wh \mu(S_{ij})$, $M_{x,i} = \msi(\msi - 1)$, and $h_R>0$ is the bandwidth. Similarly, we can estimate the variance function $V(t) = \var\{ W(t)\}= R(t,t) + \sigma_u^2$ by $\wh V(t)=\wh a_0$  where
\[
	(\wh a_0, \wh a_1) = \underset{a_0, a_1}{\argmin} \dfrac{1}{n} \sum_{i=1}^n \dfrac{1}{\msi} \sum_{j=1}^{\msi} \{L_{ij}^2 - a_0 - a_1 (S_{ij} - t)\}^2 K_{h_{V}} (S_{ij} - t).
\]
Then we estimate $\sigma^2_u$ by 
\bse
	\wh \sigma_u^2 = \dfrac{1}{|\mathcal{T}|} \int_{\mathcal{T}} \{\wh V(t,t) - \wh R(t,t)\}dt.
\ese
To estimate the functional principal components, we take a spectral decomposition of $\wh R(s,t)$ 
\bse
	\wh R(s,t) = \hbox{$\sum_{k}$} \wh \omega_k \wh \psi_k(s) \wh \psi_k(t),
\ese
which can be solved numerically by discretizing the smoothed covariance.
%It is a common practice that first a few principal components are enough for an approximation to the infinite-dimensional process. Therefore we assume $\omega_k = 0, k > q$ for some $q < \infty$. Let $\BW_i = (W_{i1}, \ldots, W_{i\msi})\trans$, 

Let $\wh \bdmu_i$, $\wh\bdpsi_{ik}$, $\wh \BPsi_i$, $\wh \BLambda$, and $\wh \BSigma_{i}$ be the estimated counterparts of  $ \bdmu_i$, $\bdpsi_{ik}$, $ \BPsi_i$, $ \BLambda$, and $ \BSigma_{i}$ using the kernel estimators described above. We adopt the PACE method of \cite{yao2005functional} to estimate the principal component score $\xi_{ik}$ by a sample version of BLUP (\ref{eq:fpc_score_blup}), i.e., 
\be\label{eq:pace}
	\wh \bdxi_{i} = (\wh \xi_{i1}, \ldots, \wh\xi_{iq})\trans= \wh \BLambda \wh \BPsi_{i}\trans \wh \BSigma_{i}^{-1} \{\BW_i - \wh \bdmu_i\}.
\ee
With (\ref{Xi_expansion}), we can recover the covariate process by
\be \label{eq:X_hat}
	\wh X_i(t) = \wh \mu(t) + \hbox{$\sum_{k=1}^q$} \wh \xi_{ik} \wh \psi_k(t), \quad t \in \mathcal{T}.
\ee

%the marginal pseudo-Gaussian log-likelihood \citep{rice2001nonparametric} is defined as 
%\[
%\wh L(q) = \sum_{i=1}^n \left\{ -\dfrac{\msi}{2} \log(2\pi) - \dfrac{1}{2} \log(det \wh \BSigma_{iq}) - \dfrac{1}{2}(\BW_i - \wh \bdmu_i)\trans \wh \BSigma_{iq}^{-1} (\BW_i - \wh \bdmu_i)\right\},
%\]

The number of components $q$ can be selected by minimizing the Akaike information criterion (AIC). There are two commonly used versions of AIC based on  a marginal log-likelihood \citep{rice2001nonparametric},
\be
AIC_{marg}(q) = \sum_{i=1}^n \left\{m_i \log(2\pi) + \log(\det \wh \BSigma_{iq}) + (\BW_i-\wh \bdmu_i)\trans \wh \BSigma_{iq}^{-1}(\BW_i - \wh \bdmu_i)\right\}+2q,
\ee
and a conditional log-likelihood \citep{li2013selecting}, 
\be\label{eq:aic}
	%q = \underset{q \geq 1}{\argmin } 
	AIC_{cond}(q) = N \log\bigg(N^{-1} \sum_{i=1}^n ||\wh \sigma_u^2 \wh \BSigma_{iq}^{-1} (\BW_i - \wh \bdmu_i)||_2^2\bigg) + N + 2nq,
\ee
where $N = \sum_{i=1}^n \msi$, $ \wh \BSigma_{iq} = \wh \BPsi_{iq}  \wh\BLambda_{q} \wh\BPsi_{iq}\trans + \wh \sigma_u^2 I$, $\wh\BLambda_{q} = \diag (\wh \omega_1, \ldots, \wh \omega_{q})$,  $\wh \BPsi_{iq} = [\wh \bdpsi_{i1}, \ldots, \wh \bdpsi_{iq}]$, and the subscript `$q$'  emphasizes the dependence on the number of FPC's.

%In terms of approximating $X_i$ by a finite number of components, $AIC$ and $AIC_{cond}$ do not differ much to influence the results. 
%
%
%
%
%
%
%
%
%
%
%
%
%
%
%
%
%
%
%
%
\vspace{-0.2cm} 
\subsection{Asynchronous Regression using Calibrated Covariates}
%Let response observations be $\BY_i = (Y_{i1}, \ldots, Y_{i\mti}) \trans$, $\BX_i = \{X_i(S_{i1}), \ldots, X_i(S_{i\msi})\}\trans$ denotes the time-varying function values at covariate time points and $\BX_{*i} = \{X_i(T_{i1}), \ldots, X_i(T_{i\mti})\}\trans$ denotes the value on response time points. Ideally, we can fit the simple regression model (\ref{ti_model}) using synchronized data $\{Y_{ij}, X_i(T_{ij})\}$, $i = 1, \ldots, n, j = 1, \ldots, \mti$. However, $X_i(T_{ij})$ is unknown, we use $\wh X_i(T_{ij})$ (\ref{eq:X_hat}) obtained via functional calibration, and estimate time-invariant coefficient as

%Interpolating predicted trajectory $\wh X_i(t)$ defined in (\ref{eq:X_hat}) on observation times of the response variables, we get calibrated values of the synchronized covariates $\wh \BX_{*i} = \{\wh X_i(T_{i1}), \ldots, \wh X_i(T_{i\mti})\}\trans$. 

%As defined above, 
{With $\wh \bdmu_{*i}$, $\wh\bdpsi_{*ik}$ and $\wh \BPsi_{*i}$  (the kernel estimates interpolated at $\BT_i$ instead of $\BS_i$),  } the empirical version of the calibrated covariate $\wt \BX_{*i}$ is $\wh \BX_{*i}= \wh\bdmu_{*i}+ \wh \BPsi_{*i} \wh \bdxi_i$, and the design matrices using calibrated covariates are $\mathbb{X}_i=(\bone, \wh\BX_{*i})$, $i=1,\ldots, n$.  %$\mathbb{X}_{ij} = \{1, \wh X_i(T_{ij})\}\trans$. 
The regression coefficients in Model (\ref{ti_model}) can be estimated by
\be \label{eqn:beta_est}
	\wh \bdbeta := (\wh \beta_0, \wh \beta_1)\trans = ( \MX \trans  \MX)^{-1} \MX \trans \BY,
\ee
where $\BY = (\BY_1\trans, \ldots, \BY_n\trans)\trans$ and $\MX = (\MX_1\trans , \ldots, \MX_{n}\trans )\trans$.

For Model (\ref{tv_model}), we estimate $\bdbeta(t):= \{\beta_0(t), \beta_1(t)\}\trans$ by using the local linear estimator of \cite{hoover1998nonparametric}, with the unobserved $X_i(T_{ij})$ replaced by the calibrated value $\wh X_i(T_{ij})$ as defined in (\ref{eq:X_hat}). That is, we estimate $\bdbeta(t)$ by $\wh \bdb_0$ from
\be \label{eqn:betat_est}
(\wh \bdb_0, \wh \bdb_1) = \underset{\bdb_0, \bdb_1}{\argmin} \sum_{i=1}^n \sum_{j=1}^{\mti} \{Y_{ij} - \mathbb{X}_{ij}\trans \bdb_0 - \mathbb{X}_{ij}\trans \bdb_1(T_{ij} - t) \}^2 K_h(T_{ij} - t),
\ee
where $\MX_{ij}=\{1, \wh X_i(T_{ij})\}\trans$ and $h>0$ is the bandwidth.
%to fit the varying-coefficient model.

 %
%
%
%
%
%
%
%
%
%
%
%
%
%
%
%
%
%
%
%
 
\section{Asymptotic Theory} \label{sec:asymptotics}
\subsection{Preliminaries}\label{sec:asymp_prelim}
For ease of exposition, we assume in the  asymptotic theory that both $X(t)$ and $Y(t)$ have been centered, such that $\mu(t)\equiv 0$ in (\ref{Xi_expansion}), $\beta_0=0$ in (\ref{ti_model}) and $\beta_0(t)=0$ in (\ref{tv_model}). 
We focus on estimating the slope parameter $\beta_1$ and the slope function $\beta_1(t)$ in Models (\ref{ti_model}) and (\ref{tv_model}), respectively; {extensions to non-centered situations are straightforward but with more
notation.} For any positive constant 
sequences $\{a_n\}$ and $\{b_n\}$, denote by $a_n \prec b_n$ if $a_n/ b_n \to 0$ as $n\to \infty$.

Recall $\BX_{*i}$ is the unobserved covariate vector synchronized with the response $\BY_i$, $\wt \BX_{*i} = \BPsi_{*i} \wt \bdxi_i$ is the best linear unbiased predictor of $\BX_{*i}$ as defined in (\ref{eq:X_tilde_star}), and $\wh \BX_{*i}$ is the empirical version of $\wt \BX_{*i}$ replacing the unknown functions with their kernel estimators.
%where $\wt \bdxi_i$ is defined in (\ref{eq:fpc_score_blup}), $\BPsi_{*i} = (\bdpsi_{*i1}, \ldots, \bdpsi_{*iq})$, and $\bdpsi_{*ik} = \{\psi_k(T_{i1}), \ldots, \psi_k(T_{i\mti})\}\trans$, $k=1,\ldots, q$. 
Let $\bdepsilon_i=\{\epsilon_i(T_{i1}), \ldots, \epsilon_i(T_{i\mti})\}\trans$ be the vector of measurement error as defined in model (\ref{ti_model}) or (\ref{tv_model}), with the covariance matrix $\BOmega_i = \{ \Omega(T_{ij}, T_{ij'})\}_{j, j'=1}^{\mti}$.%\E(\bdepsilon_i \bdepsilon_i\trans \mid \msi, \mti, T_i, S_i)$.

We assume that the numbers of observations $(\msi,  \mti)$  are random with $P(2 \leq \msi, \mti \leq M) = 1$ for a constant $M < \infty$. Given $\msi$, $S_{ij}$ are iid copies of the random variable $S$ with a density $f_S(\cdot)$; and given $\mti$, $T_{ij}$ are iid copies of the random variable $T$ with a density $f_T(\cdot)$.  Both $f_T(\cdot)$ and $f_S(\cdot)$ are strictly greater than 0, with bounded derivatives on $\CT$. %Denote $\BT_i=(T_{i1}, \ldots, T_{i \mti})\trans$ and $\BS_i=(S_{i1},\ldots, S_{i\msi})\trans$, and 
Assume that $\{X_i(t), \epsilon_i(t), \msi, \mti,  \BT_{i}, \BS_{i}\}$ are iid tuples across $i$. In addition, we assume the following conditions hold. %\xc{this paragraph states some conditions that does not list in below, although $\mu$ is assumed zero do we need to give $h_{\mu}$ condition?\\add a Remark4: }

\begin{enumerate}[label=(C.\arabic*)]
%     \item \label{sparse_assum} The number of observations $(\msi,  \mti)$ are iid realizations of a random variable bounded by $M < \infty$ almost surely, i.e., $P(2 \leq \msi \leq M) = 1$ and $P(2 \leq \mti \leq M) = 1$
%     \item \label{f_assum} 
%     Given $\msi$, $S_{ij}$ are iid copies of the random variable $S$; given $\mti$, $T_{ij}$ are iid copies of the random variable $T$. The density $f_T(\cdot)$ of $T$ and density $f_S(\cdot)$ of $S$ are both bounded from below and above. Further, $f_T$ and $f_S$ are both differentiable with bounded derivative.
	%$$ 0 < \underset{t \in \mathcal{T}}{\min} f_T(t) \leq \underset{t \in \mathcal{T}}{\max} f_T(t) < \infty, \quad 0 < \underset{t \in \mathcal{T}}{\min} f_S(t) \leq \underset{t \in \mathcal{T}}{\max} f_S(t) < \infty, $$
	%with density functions $f_{m,x}(\cdot)$ and $f_{m,y}(\cdot)$ respectively.
	%and the second derivative $f_T^{(2)}(\cdot)$ and $f_S^{(2)}(\cdot)$  are bounded.
	\item \label{kernel_assum} The kernel function $K(\cdot)$ is a symmetric probability density function on $[-1, 1]$ with
	\[
	\sigma_K^2 := \int_{-1}^1 u^2 K(u) du < \infty, \quad \nu_0 := \int_{-1}^1 K^2(u) du < \infty.
	\]
% 	\item \label{indep_assum} The underlying processes $X_i$, the number of observations $(m_{x,i}, m_{y,i})$ and the observation time points $(T_{i}, S_{i})$ are mutually independent. The measurement errors $\epsilon_{i}(t), t \in \mathcal{T}$ are identically distributed and are independent of the random element $\{X_i, m_{x,i}, m_{y,i}, T_{i}, S_{i}\}$.
% 	\item \label{mu_assum} $\mu(t)$ is twice differentiable and the second derivative $\mu^{(2)}(t)$ is bounded on $\mathcal{T}$.
	\item \label{cov_cond} All eigenfunctions $\psi_k(t)$, $k=1,\ldots, q$, are twice  differentiable, and the second derivatives $\psi_k^{''}(t)$ are uniformly continuous on $\CT$.
	%All second-order partial derivatives of $R(s,t)$ exist and are bounded on $\mathcal{T}^2$.
%	\item \label{mm_cond} $\E(|U_{ij}|^{2 \lambda_R}) < \infty$ and $\E\left\{\underset{t \in \mathcal{T}}{\sup}|X_i(t)|^{2 \lambda_R}\right\} < \infty$ for some $\lambda_R > 2$; $h_R \rightarrow 0$ and $(h_R^4 + h_R^3 + h_R^2)^{-1} (\log n/n)^{1-2/\lambda_R} \rightarrow 0$ as $n \rightarrow \infty$.
%	\item \label{mm_cond2} $\E(|U_{ij}|^{2 \lambda_V}) < \infty$ and $\E\left\{ \underset{t \in \mathcal{T}}{\sup}|X_i(t)|^{2 \lambda_V}\right\} < \infty$ for some $\lambda_V > 2$; $h_V \rightarrow 0$ and $(h_V^2 + h_V)^{-1} (\log n / n)^{1-2/\lambda_V} \rightarrow 0$  as $n \rightarrow \infty$.
	\item \label{mm_cond} There exists a constant $C>4$ such that $\E(|U_{ij}|^C) +\E\left\{\underset{t \in \mathcal{T}}{\sup}|X_i(t)|^{C}\right\} < \infty$.
	\item \label{bw_cond} Assume $h_R \to 0$, $h_V \rightarrow 0$ as $n \rightarrow \infty$, %$(\log n)n^{-1/2} \prec h_R \prec n^{-1/4}$ and $(\log n) n^{-3/4} \prec h_V \prec n^{-1/4}$.
	$(\log n/n)^{1/3} \prec h_R \prec n^{-1/4}$ and $(\log n) n^{-3/4} \prec h_V \prec n^{-1/4}$.
	%and $h_V \rightarrow 0$ as $n \rightarrow \infty$ and satisfies: $\log n/\sqrt{n} \prec h_V \prec n^{-1/4}$.
\end{enumerate}

\noindent {\bf Remark 1.} These conditions  are common in functional data analysis. Under them,  \cite{li2010uniform} proved that the FPCA estimators possess the following uniform convergence properties 
\bse
	\underset{s,t \in \mathcal{T}}{\sup}|\wh R(s,t) - R(s,t)| &=& O\{h_R^2 + \sqrt{\log n/(nh_R^2)}\} \quad a.s., \\
	\wh \sigma_u^2 - \sigma_u^2 &=& O\{h_R^2 + \sqrt{\log n/(nh_R)} + h_V^2 + \log n/(nh_V)\} \quad a.s., \\
	\underset{t \in \mathcal{T}}{\sup}|\wh \psi_k(t) - \psi_k(t)| &=& O\{h_R^2 + \sqrt{\log n /(nh_R)} \} \quad a.s.,\\
	\wh \omega_k - \omega_k &=& O(\sqrt{\log n/n}) \quad a.s. \quad k=1,\ldots, q.
\ese
Recall $\wt X_i(t)$ defined in (\ref{eq:X_tilde}) is  the BLUP for $X_i(t)$ and $\wh X_i(t)$ in (\ref{eq:X_hat}) is its empirical counterpart. By straightforward calculations,
\be \label{Xhat_Xtilde}
	 \underset{t \in \mathcal{T}}{\sup} |\wh X_i(t) - \wt X_i(t)| 
	%& = & \underset{t \in \mathcal{T}}{\sup} \abs{\sum_{k=1}^{q} \wh \omega_k \wh \psi_k(t) \wh \bdpsi_{is,k} \trans \wh \BSigma_i^{-1} \BW_i - \sum_{k=1}^{q} \omega_k \psi_k(t) \bdpsi_{is,k}\trans \BSigma_i^{-1} \BW_i}\\ \nonumber
	%& \leq & \sum_{k=1}^{q} \underset{t \in \mathcal{T}}{\sup} |\wh \omega_k \wh \psi_k(t) - \omega_k \psi_k(t) | \abs{\wh \bdpsi_{is,k} \trans \wh \BSigma_i^{-1} \BW_i}  + \sum_{k=1}^q \underset{t \in \mathcal{T}}{\sup} |\omega_k\psi_k(t)| \abs{(\wh \bdpsi_{is,k} \trans \wh \BSigma_i^{-1} - \bdpsi_{is,k} \trans \BSigma_i^{-1}) \BW_i} \\ 
	= O\{ h_R^2 + \sqrt{\log n/(nh_R)} + h_V^2 + \log n/(nh_V)\} \quad a.s.
\ee
%
%where $\wh \BSigma_i = \wh \BPsi_{i} \wh \BLambda \wh \BPsi_{i} \trans + \wh \sigma^2 \BI$. 
It is well-known in the measurement error literature \citep{Carroll2006}, replacing $X_i(t)$ with the calibrated value $\E\{ X_i(t) \mid \BW_i\}$ will result in consistent but less efficient estimators. Equation (\ref{Xhat_Xtilde}) shows that our functional calibration $\wh X_i(t)$ uniformly converges to the BLUP $\wt X_i(t)$, and hence our estimators in (\ref{eqn:beta_est}) and (\ref{eqn:betat_est}) are consistent; however, the derivation of their asymptotic distributions needs much involved analysis.

%
%
%
%
%
%
%
%
%
%
%
%
%
%
 
%\subsection{Regression with Time-invariant Coefficients}
\subsection{Asymptotic Properties of FCAR Estimator for the Time-Invariant Regression Model}\label{sec:aymp_ti}

%Assuming that both $X_i(t)$ and $Y_i$ are centered with mean zero, and the slope estimator for the time-invariant regression model (\ref{ti_model}) can be simplified as  
%\bse
%	\wh \beta_1 = \dfrac{\sum_{i=1}^n \sum_{j=1}^{\mti} \wh X_i(T_{ij})Y_{ij}}{\sum_{i=1}^n \sum_{j=1}^{\mti} \wh X_i^2(T_{ij})}, \quad  \hbox{where }\wh X_i(T_{ij}) = \sum_{k=1}^q \wh \xi_{ik} \wh \psi_k(T_{ij}).
%\ese
%Under the simplified setting that both $X_i(t)$ and $Y_i$ are centered with mean zero, 
The following theorem establishes the asymptotic property of the coefficient estimator (\ref{eqn:beta_est}) under the time-invariant regression model (\ref{ti_model}). 
\begin{theorem} \label{thm:beta} 
%(Time-invariant regression model) 
Under the assumptions above, the estimated slope parameter for model (\ref{ti_model}) has the following asymptotic distribution
%there exists a variance function $\gamma^2(\beta_1) =  \{\wt \gamma^2(\beta_1) + \gamma_0^2\beta_1^2\}/\gamma_x^2$, such that 
\bse
	\sqrt{n}(\wh \beta_1 - \beta_1) \cid \Normal\{ 0, (\gamma_1+\beta_1^2 \gamma_2)/\gamma_x^2 \},
\ese
where %$\gamma_x =\E (\wt \BX_{*i}\trans \wt \BX_{*i})= \E \left\{\tr\left(\BPsit_{i} \BLambda \BPsi_{i} \trans \BSigma_{i}^{-1} \BPsi_i \BLambda \BPsit_i \trans \right) \right\}$,
\be \label{eq:ti_asymp_var}
	\gamma_x &=& \E (\wt \BX_{*i}\trans \wt \BX_{*i})= \E \left\{\tr\left(\BPsit_{i} \BLambda \BPsi_{i} \trans \BSigma_{i}^{-1} \BPsi_i \BLambda \BPsit_i \trans \right) \right\}, \nonumber \\ 
	\gamma_1&= & \E \bigg\{ \tr(\BPsi_{*i} \BLambda \BPsi_i\trans \BSigma_i^{-1} \BPsi_i \BLambda \BPsi_{*i} \trans \BOmega_i)\bigg\}, \nonumber\\
	\gamma_2 &=& \var \bigg[ \sum_{j=1}^\mti \wt X_{i}(T_{ij}) \{ X_{i}(T_{ij})-\wt X_{i}(T_{ij})\} + \dfrac{1}{M_{x,i}} \sum_{j \neq j'} u^*_{i, jj'} \CA(S_{ij}, S_{ij'}) \bigg], 
\ee
$u_{i,jj'}^*= W_{ij} W_{ij'} - R(S_{ij}, S_{ij'})$,  $\CA(s_1, s_2)$ is defined in Lemma 2 in the Appendix, and the  expectations  are taken over $(\BX_i, \bdepsilon_i, \msi, \mti, \BT_i, \BS_i)$.
\end{theorem}
\noindent {\bf Remark 2.} Under the special case that the BLUP $\wt X_i(t)$ is also the conditional mean $\E\{ X_i(t) | \BW_i\}$, for example when $\bdxi_i$ and $\BU_i$ are jointly Gaussian, $\{ \wt X_i(t)- X_i(t), \ t\in \CT\}$ is uncorrelated with any function of $\BW_i$. One can show, under such an circumstance, $\gamma_2 =\gamma_{21} +\gamma_{22}$, where $\gamma_{21}=\var[ \sum_{j=1}^\mti \wt X_{i}(T_{ij}) \{ X_{i}(T_{ij})-\wt X_{i}(T_{ij})\} ]$ and $\gamma_{22}= \var\{ M_{x,i}^{-1} \sum_{j \neq j'} u^*_{i, jj'} \CA(S_{ij}, S_{i'j'})\}$.
Under the additional Gaussian assumption on $\bdxi_i$ and $\BU_i$, we can also obtain
$$
	 \gamma_{21} = \E \left[ \tr\left\{ \BPsit_{i} \BLambda \BPsi_{i} \trans \BSigma_{i}^{-1} \BPsi_i \BLambda \BPsit_i \trans  \BPsit_i \BLambda (\BI - \BPsi_i\trans \BSigma_{i}^{-1} \BPsi_i \BLambda ) \BPsit_i\trans \right\} \right].
	%\Pi(\beta_1) = \beta_1^2 \BPsit_i \BLambda_K (\BI - \BPsi_i\trans \BSigma_{iq}^{-1} \BPsi_i \BLambda_K ) \BPsit_i\trans + \BOmega_i.
%
$$

As in the classic regression calibration literature \citep{Carroll2006}, one can define $\wt \bdbeta$ to be the counterpart of $\wh \bdbeta$ in (\ref{eqn:beta_est}) replacing $\wh X_i(t)$ with $\wt X_i(t)$, as if all the functional and scalar parameters in (\ref{eq:disc_obs}) and (\ref{Xi_expansion}) are known, then $(\gamma_1 +\beta_1^2 \gamma_{21})/ \gamma_x^2$ is the asymptotic variance of $\wt \beta_1$. In our problem, $\beta_1^2 \gamma_{22}/ \gamma_x^2$ is the additional variation caused by the FPCA estimation errors, i.e. those caused by substituting $\mu(t)$, $\psi_k(t)$, $\omega_k$ and $\sigma_u^2$ with their functional estimators described in Section \ref{sec:methods:functional_calibration}. 
.

{While it is tempting to treat the calibrated values $\wh \BX_{*i}$ as the truth  and use the naive standard error for linear regression to  infer $\beta_1$,} the decomposition of the asymptotic variance in Theorem \ref{thm:beta} suggests that this approach ignores the extra variations {caused by calibration of the covariate values as well as estimation errors from FPCA.}  As a result, the naive approach leads to an underestimated variation, a low coverage rate in confidence intervals and illegitimate inferences. We  recommend estimating the standard error of $\wh \bdbeta$ using bootstrap, where we resample the subjects and repeat the FPCA procedure to the bootstrap samples to properly account for {these extra variations.}

\subsection{Asymptotic Properties of FCAR Estimator for the Time-Varying Regression Model}\label{sec:asymp_tv}
Again we  assume both $X(t) $ and $Y(t)$ are centered  so that $\beta_0(t)=0$ and we can focus on estimating $\beta_1(t)$ in Model (\ref{tv_model}). We also make the additional assumptions.

\begin{enumerate}[label=(C.\arabic*)]
\setcounter{enumi}{4}
\item The slope function $\beta_1(t)$ is twice continuously differentiable on $\CT$.
\item The kernel function in fitting the time-varying regression model in (\ref{eqn:betat_est}) is Lipschitz continuous and satisfies \ref{kernel_assum}.
\item\label{cond:tv_bandwidth} The bandwidth $h$ in (\ref{eqn:betat_est}) satisfies $h_R/ h \to 0$, $\log(n) h^5/h_R\to 0$, $nh^7 \to 0$ and $nh \to\infty$. 
\end{enumerate}
%\boxit{Need more assumptions for the smoothness of $\beta_1(t)$ and the bandwidth $h$}

%
%
%
%
%
%
%
%

\begin{theorem} \label{thm:beta_t}
Under the framework outlined in Section \ref{sec:asymp_prelim} and assumptions \ref{kernel_assum}---\ref{cond:tv_bandwidth}, the estimated slope function for model (\ref{tv_model}) has the following asymptotic distribution
\[
	\sqrt{nh} \Gamma_0(t)\{\wh \beta_1(t) - \beta_1(t) - \dfrac{1}{2} \beta_1^{(2)}(t) \sigma_K^2 h^2\} \cid \Normal[ 0,  \Gamma_1(t) + \beta_1^2(t) \{\Gamma_2(t)+ \Gamma_3(t)\} ],
\]
for any $t \in \mathcal{T}$, where $\bar m_y = \E(\mti)$, $\Gamma_0(t) = \bar m_y f_T(t) \Gamma_x(t)$,
\ben\label{eq:beta_t_asymp_var}
	\Gamma_x(t) &=& \var\{ \wt X_i(t)\} = \bdpsi\trans(t) \BLambda \E(\BPsi_i\trans \BSigma_i^{-1} \BPsi_i) \BLambda \bdpsi(t), \nonumber \\
	%
%	A(t) &=& \bar m_y f_T(t) \bdpsi(t)\trans \BLambda \E\{\BPsi_i\trans \BSigma_i^{-1} \BPsi_i\} \BLambda \bdpsi(t),  \nonumber \\
%	%\CB(t) &=& \dfrac{1}{2} \beta_1^{(2)}(t) \bar m_y f_T(t) \bdpsi(t)\trans \BLambda \E\{\BPsi_i\trans \BSigma_i^{-1} \BPsi_i\} \BLambda \bdpsi(t) \sigma_K^2,\\
%	%
%	\Gamma(t) &=& \bar m_y g(t) f_T(t) \nu_0, \nonumber \\
%	g(t) &=& \E\left[\wt X_i^2(T_{ij}) \{\beta_1(T_{ij}) X_i(T_{ij}) - \beta_1(T_{ij}) \wt X_i(T_{ij}) + \epsilon_{i}(T_{ij})\}^2 | T_{ij} = t\right],
%	
	%\Gamma_0(t) &=& \bar m_y f_T(t) \Gamma_x(t),  \nonumber\\
	%
	\Gamma_1(t)&=&  \bar m_y \Gamma_x(t) \Omega(t,t) f_T(t) \nu_0, \nonumber \\
	\Gamma_2(t)&=& \bar m_y \E[ \wt X^2(t) \{ X(t)- \wt X(t)\}^2] \nu_0 , \nonumber\\
	%\Gamma_3(t)&=&\bar m_y^2  f_S(t) \E(M_{x,i}^{-1})  \nu_0 \int \Pi(s, t) \CQ(s, t) f_S(s) ds , 
	\Gamma_3(t) &=& \bar m_y^2 f_S(t)\nu_0 \bigg[ \E(M_{x,i}^{-1}) \int \Pi(t, s_2,s_2) \CQ^2(s_2,t) f_S(s_2) ds_2 \nonumber \\
&& + \E\{M_{x,i}^{-1}(m_{x,i}-2)\}\int \Pi(t,s_2,s_3) \CQ(s_2,t) \CQ(s_3,t) f_S(s_2)f_S(s_3) ds_2ds_3\bigg],\nonumber
\een
%
%$\CQ(s,t)=\bdpsi\trans(t) \BLambda \E( \BPsi_i\trans \BSigma_i^{-1} \BPsi_i  ) \bdpsi(s)/ f_S(s)$, $\Pi(s, t)= \E\{ X^2(s) X^2(t)\} + \sigma_u^2 R(s,s) + \sigma_u^2 R(t,t)+ \sigma_u^4$. 
$\CQ(s, t) = \bdpsi\trans(t) \BLambda \E( \BPsi_i\trans \BSigma_i^{-1} \BPsi_i  ) \bdpsi(s)f_T(t)/ \{f_S(s)f_S(t)\}$, 
%$\Pi(s_1,s_2,s_3) = \E\{X^2(s_1)X(s_2)X(s_3)\} + \E\{X(s_2)X(s_3)\}\sigma_u^2 - \E\{X(s_3)X(s_1)\}R(s_2,s_1) - \E\{X(s_2)X(s_1)\}R(s_3,s_1) + R(s_2,s_1)R(s_3,s_1) + I(s_2=s_3) \E\{X^2(s_1)\}\sigma_u^2 + \sigma_u^4 I(s_2=s_3)$.
$\Pi(s_1,s_2,s_3) = \E\{X^2(s_1)X(s_2)X(s_3)\} + R(s_2,s_3) \sigma_u^2 - R(s_1,s_2)R(s_1,s_3) + I(s_2=s_3) \{ R(s_1, s_1) \sigma_u^2 + \sigma_u^4 \}$, and
 the expectations are taken over $(\msi, \mti, \BT_i, \BS_i, \BX_i, \bdepsilon_i)$.
\end{theorem}
\noindent {\bf Remark 3.} 
 Theorem \ref{thm:beta_t} suggests that our estimator enjoys the optimal convergence rate in varying coefficient models as established in \cite{Cai2000jasa}, which is much faster than those for the competing method of \cite{cao2015regression} and \cite{csenturk2010functional}. 
Analogous to Theorem \ref{thm:beta},  $\Gamma_0^{-2}(t) \{\Gamma_1(t) + \beta_1^2(t) \Gamma_2(t)\}$ is the asymptotic variance of $\wt \beta_1(t)$, obtained by using $\wt \BX_{*i}$ as the predictors in the varying coefficient model (\ref{tv_model}), and $\Gamma_0^{-2}(t) \beta_1^2(t) \Gamma_3(t)$ is the extra variation caused by the FPCA errors. We therefore recommend to make inference on $\bdbeta(t)$ using a bootstrap procedure that accounts for the FPCA estimation error as described in Remark 2. Also Assumptions \ref{bw_cond} and \ref{cond:tv_bandwidth} require undersmoothing in the FPCA procedure;  we need $h_R/ h \to 0$ so that the biases caused by FPCA estimation are asymptotically negligible compared with the smoothing bias in varying coefficient models.  

%Note that both $\Gamma(t)$ and $\Gamma^{\dag}(t)$ depend on the true coefficient function $\beta_1(t)$. If we assume $nh^5 \rightarrow 0$ and bounded $\beta_1^{(2)}(t)$, the bias caused by linear kernel estimator of coefficient converges to zero as $n \rightarrow \infty$. Thus, Theorem~\ref{thm:beta_t} shows that our proposed time-varying estimator is consistent and asymptotic normal. Similar to  the explanation in Theorem~\ref{thm:beta}, $\Gamma(t)$ is part of the asymptotic variance induced by time-varying coefficient regression using $\wt X_i(T_{ij})$ as inputs, while $\gamma_0^2(t)$ is from imputation error in the first step of functional calibration.

%\noindent {\bf Remark 4.} centered variables in regression means $X_i(T_{ij}) - \dfrac{1}{N}\sum_{i = 1}^n\sum_{j =i}^{m_i} X_i(T_{ij}) $, but we assume $\mu(t) = 0$ here in the proof.

%
%
%
%
%
%
%
%
%
%
%
%
%
%
%
%
%
%
%
%
 
\section{Simulation Studies} \label{sec:simulation}
We conduct simulations to examine the finite sample performances of  the time-invariant regression model (\ref{ti_model}) and time-varying coefficient model (\ref{tv_model}), and compare them with those of various exiting methods.

%We will compare the numerical performance of our proposed FCAR method with other existing methods: kernel weighted  method \citep{cao2015regression} denoted as WK, functional varying coefficients model \citep{csenturk2013modeling} denoted as FVCM, and the last observation carried forward method denoted as LOCF. 
%\subsection{Functional Calibration for Asynchronous Regression (FCAR)} \label{sec:simulation:fcar}
\subsection{Simulation 1: FCAR for time-invariant coefficient model} \label{sec:simulation:fcar}

Let the time domain be $\CT=[0,10]$, $X_i(t)$ be iid copies of a stochastic process described by model (\ref{Xi_expansion}) with $q=3$ principal components, and $\bdxi_i \sim \Normal\{ \pmb{0}, \diag(4, 2, 1)\}$. Set $n = 200$ and generate $Y_i(t)$ from Model (\ref{ti_model}) with $\beta_0 = 1$ and $\beta_1 = 2$. Suppose there are $m=5$ discrete observations on $X_i(t)$ and $Y_i(t)$, respectively, where $\{S_{i1}, \ldots, S_{im}\}$ and $ \{T_{i1},\ldots, T_{im}\}$ are generated independently from a uniform distribution on $\CT$. Error-contaminated discrete observations $\BW_i$ are generated from Model (\ref{eq:disc_obs}) with $U_{ij}\sim \Normal(0,1)$.
We consider two settings for the mean and eigenfunctions of $X_i(t)$: 

\noindent     \textit{Setting I}: $\mu(t) = t + \sin(t)$, $\psi_{k}(t) = (1/\sqrt{5}) \sin(\pi k t/10)$, $t \in \CT$, $k=1,2, 3$; \\
     \textit{Setting \II}: $\mu(t) = \sin(t)$, 
$\psi_{1}(t) = \sqrt{10}$, $\psi_2(t) = \sqrt{5} \sin(2\pi t/10)$, $\psi_3(t) = \sqrt{5}\cos(2\pi t/10)$.

%Let time points $S_{ij}$ and $T_{ij}$ are independent and iid in time domain, following $Uniform(0, 10)$. The simulation model is
%\bse
%Y_i(T_{ij}) &=& \beta_0 + \beta_1 X_i(T_{ij}) + \epsilon_i(T_{ij}), \quad X_i(T_{ij}) = \mu(T_{ij}) + \sum_{k=1}^{3} \xi_{ik} \psi_{k}(T_{ij}), \\
%W_i(S_{ij}) &=& \mu(S_{ij}) + \sum_{k=1}^{3} \xi_{ik} \psi_{k}(S_{ij}) + u_{ij},
%\ese
%where $u_{ij}$ are iid measurement errors following standard normal distribution, 
\noindent We generate residual $\epsilon_i(t)$ from a zero-mean Gaussian process with covariance function $\Omega(s, t) = \cov\{\epsilon_i(s), \epsilon_{i}(t)\}$, and consider two different covariance structures: 1) independent  \xc{(IE)} with $\Omega(s, t) = 1.5I(s=t)$ and 2) dependent  \xc{(DE)} with $\Omega(s, t) = 2^{-|t-s|/5}$. \yl{As an  ideal case,   we also  consider a measurement-error free (MEF) scenario under the DE structure, where $X_{ij}$'s in  (\ref{eq:disc_obs}) are correctly observed and the covariate measurement error $U_{ij}=0$}.

For each setting and each error correlation structure, we simulate 200 data sets and apply the proposed FCAR method to each simulated data set. Specifically, FPCA is performed using the \texttt{fdapace} package of R with its built-in bandwidth selector and $q$ is selected by the marginal likelihood AIC. In Table \ref{tbl:sim:fcar}, we summarize the performance of $\wh \beta_1$ under both settings and \xc{all three measurement error structures (IE, DE and MEF)}. The criteria include the bias, standard deviation, mean of the naive standard error pretending the calibrated values are the true covariates, coverage rate of a 95\% confidence interval using the naive SE, mean of the bootstrap standard error, and coverage rate of a 95\% confidence interval using the bootstrap SE. The results on $\wh \beta_0$ are similar but less interesting and hence relegated to Appendix C. It appears that the bias of our estimator is much smaller than the standard deviation, corroborating Theorem \ref{thm:beta} that $\wh\beta_1$ is asymptotically unbiased. The results also support Remark 2 that the naive standard error estimator underestimates the standard error and results in confidence intervals with lower than nominal coverage rates. In contrast, bootstrap standard errors capture the extra variations caused by calibrating the covariate value and FPCA estimation errors, and as a result the confidence intervals based on bootstrap standard errors yield coverage rates close to the nominal ones. As noted in Remark 2, we perform FPCA to each bootstrap sample, and Table \ref{tbl:sim:fcar} is based on 500 bootstrap samples.

\begin{table}[ht]
	\centering
	\begin{tabular}{cccc|ccc}
		\hline 
		\hline 
		 & \multicolumn{3}{c|}{Setting I} & \multicolumn{3}{|c}{Setting \II} \\
Error type & IE  & DE & MEF & IE  & DE & MEF \\
		 \hline
Bias & 0.007 & 0.004 & -0.002 & -0.008 & -0.013 & 0.025 \\ 
SD & 0.028 & 0.029 & 0.017 & 0.127 & 0.122 & 0.060 \\ 
Naive SE & 0.019 & 0.017 & 0.012 & 0.064 & 0.058 & 0.034 \\ 
Naive CP & 0.830 & 0.770 & 0.820 & 0.670 & 0.640 & 0.725 \\ 
Bootstrap SE & 0.030 & 0.030 & 0.019 & 0.117 & 0.119 & 0.064 \\ 
Bootstrap CP & 0.955 & 0.950 & 0.965 & 0.925 & 0.930 & 0.940 \\ 
%$s.e.(\wh \beta_0)$ & 0.1795 & 0.187 & 0.0892 & 0.0999 \\ 
%$\mathrm{CP}(\wh \beta_0)$ & 0.965 & 0.935 & 0.95 & 0.95 \\ 
\hline
	\end{tabular}
\vspace*{18pt}
\vspace{0.5cm}
\caption[Performance of the proposed FCAR method under different scenarios.]{Simulation 1: performance of $\wh \beta_1$ under the proposed FCAR method under Settings 1 and 2. SD: standard deviation; Naive SE: mean of the naive standard error; Naive CP:  coverage rate of a 95\% confidence interval using the naive SE;  Bootstrap SE: mean of the bootstrap standard error; Bootstrap CP:  coverage rate of a 95\% confidence interval using the bootstrap SE; IE: independent errors; DE: dependent errors; MEF: model-error free with dependent errors.
%different scenarios, including bias, standard deviation, mean of the naive standard error, coverage rate of a 95\% confidence interval using the naive SE, mean of the bootstrap standard error, and coverage rate of a 95\% confidence interval using the bootstrap SE. 
}
\label{tbl:sim:fcar}
\end{table}

Table \ref{tbl:sim:compare} compares the proposed FCAR method with the kernel weighted (KW) method \citep{cao2015regression} on biases, Monte Carlo standard deviations, and the average estimated standard errors.
%; \yhl{we have omitted a comparison with the naive LOCF method because it is  biased and tends to perform even more poorly than  KW \citep{Molenberghs2004biostatistics}.}
%As it is well known that the naive LOCF method is biased and tends to perform more poorlythan  KW, we have omitted a comparison with LOCF. However, these results are available upon request.} 
For the KW method, we use the function \texttt{asynchTI} from the R package \texttt{AsynchLong} \citep{cao2015regression}, which  provides a built-in standard error estimator. For FCAR method, the standard error refers to the boostrap standard error in Table~\ref{tbl:sim:fcar}.  
%For LOCF, we impute the unobserved covariate value at $T_{ij}$ by its nearest observation prior to $T_{ij}$. 
%The bootstrap functional data procedure is not appropriate for LOCF, because it  assumes covariate stops changing after a certain point, {which does not conform to a functional data context; instead, we use the naive standard error for the LOCF coefficient estimates.} 
%{\bf There is a \texttt{asynchLV} function for LOCF in the \texttt{AsynchLong} package, with built-in standard error.}
\yl{It is noteworthy that, under the  IE and DE  covariance structures, the magnitude of the KW biases still dominates that of the standard errors, yielding  confidence intervals with a coverage rate close to 0. In contrast,  the proposed FCAR estimator incurs negligible biases and produces confidence intervals with a coverage rate close to the nominal level. The coverage rate of the KW confidence intervals  improves much  under the ideal MEF  (no measurement errors) scenario, but is still  lower than that of the FCAR confidence intervals. For the coefficient estimation, FCAR and KW are on par in  computational intensity, taking 16.08 and 14.76 seconds respectively;
the calculation of standard errors for FCAR takes 1.96 minutes in 10-cores parallel for each data set on average,  a bit more than 0.13 seconds taken by KW. This is reasonable, as  FCAR needs a bootstrap procedure to compute standard errors,  while KW does not.} \yl{Including an additional setting of $m_i=15$ per subject under the DE structure, we investigate the impact of the sparsity level on the performance and find that the performance is fairly robust; see Appendix C.}

\begin{table}[ht]
	\centering
	\begin{tabular}{c|ccc|cc|cc}
		\hline 
		\hline 
		& & \multicolumn{2}{c|}{IE} & \multicolumn{2}{|c}{DE} & \multicolumn{2}{|c}{MEF}\\
		& & FCAR & KW & FCAR & KW & FCAR & KW\\
		\hline 
Setting I & Bias & 0.007 & -0.225 & 0.004 & -0.213 & -0.002 & -0.024 \\ 
 & SD & 0.028 & 0.067 & 0.029 & 0.067 & 0.017 & 0.045 \\ 
 & SE & 0.030 & 0.049 & 0.030 & 0.051 & 0.019 & 0.032 \\ 
 & CP & 0.955 & 0.060 & 0.950 & 0.065 & 0.965 & 0.830 \\ 
Setting \II & Bias & -0.008 & -0.978 & -0.013 & -0.978 & 0.025 & -0.060 \\ 
 & SD & 0.127 & 0.097 & 0.122 & 0.092 & 0.060 & 0.108 \\ 
 & SE & 0.117 & 0.076 & 0.119 & 0.077 & 0.064 & 0.069 \\ 
 & CP & 0.925 & 0.000 & 0.930 & 0.000 & 0.940 & 0.735 \\  
\hline\hline
	\end{tabular}
	\vspace{1cm}
\caption{Simulation 1: comparison of $\wh \beta_1$ using the proposed FCAR method with the kernel weighted (KW) method of \citep{cao2015regression} on bias, standard deviation (SD), mean of standard error (SE) and coverage rate of a 95\% confidence interval using standard error (CP) under two settings and three error structures \xc{(IE: independent residuals; DE: dependent residuals; MEF: measurement error free with dependent residuals)}}
\label{tbl:sim:compare}
\end{table}

%\begin{table}[ht]
%	\centering
%	\caption[Different methods under Setting \II.]{Different methods under Setting \II: mean value (sample standard deviation) in 200 simulations.}
%\begin{tabular}{cccc|ccc}
%		\hline 
%		\hline 
%		& \multicolumn{3}{c|}{Independent error} & \multicolumn{3}{|c}{Dependent error} \\
%		& FCAR & KW & LOCF & FCAR & KW & LOCF \\
%		\hline 
%SE & 0.024(0.026) & 1.01(0.194) & 1.872(0.196) & 0.024(0.023) & 1.005(0.184) & 1.9(0.191) \\ 
%PE & 0.113(0.161) & 6.767(1.297) & 12.93(1.34) & 0.104(0.134) & 6.757(1.251) & 13.099(1.292) \\ 
%% $\Delta \beta_0$ & 0.007(0.086) & 0.166(0.126) & 0.124(0.098) & 0.008(0.093) & 0.162(0.117) & 0.13(0.111) \\ 
%$\Delta \beta_1$ & -0.008(0.127) & -0.978(0.097) & -1.357(0.07) & -0.013(0.122) & -0.978(0.092) & -1.366(0.068) \\ 
%%$s.e.(\wh \beta_0)$ & 0.089 & 0.121 & 0.08 & 0.1 & 0.126 & 0.076 \\ 
%%$\mathrm{CP}(\wh \beta_0)$ & 0.96 & 0.725 & 0.635 & 0.965 & 0.76 & 0.585 \\ 
%$s.e.(\wh \beta_1)$ & 0.117 & 0.076 & 0.054 & 0.119 & 0.077 & 0.051 \\ 
%$\mathrm{CP}(\wh \beta_1)$ & 0.93 & 0 & 0 & 0.955 & 0 & 0 \\  
%		\hline
%	\end{tabular}
%	\label{tbl:sim:compare2}
%\end{table}
%
%
%
%
%
%
%
%
%
%
%
%
%
%
%

\subsection{Simulation 2: FCAR for time-varying coefficient model} \label{sec:simulation:fcart}
We simulate data from the time-varying coefficients regression model (\ref{tv_model}). As in Simulation 1, we set the time domain to be $\CT= [0, 10]$ and simulate $n = 200$ subjects with $m_i = 5$ repeated measures on $X_i(t)$ and $Y_i(t)$ allowing the measuring time points to be asynchronous between $X$ and $Y$.  We simulate $X(t)$ using the Karhunen-Lo\`eve expansion (\ref{Xi_expansion}), with mean function $\mu(t) = t + \sin(t)$, $q=3$, $\bdxi_i \sim \Normal\{ \pmb{0}, \diag(4, 2, 1)\}$ and $ \psi_{k}(t) = (1/\sqrt{5}) \sin(\pi k t/10)$, $k =1, 2, 3$. 
%The simulation model is 
%\bse
%Y_i(T_{ij}) &=& \beta_0(T_{ij}) + \beta_1(T_{ij}) X_i(T_{ij}) + \epsilon_i(T_{ij}), \quad X_i(T_{ij}) = \mu(T_{ij}) + \sum_{k=1}^{3} \xi_{ik} \psi_{k}(T_{ij}), \\
%W_i(S_{ij}) &=& \mu(S_{ij}) + \sum_{k=1}^{3} \xi_{ik} \psi_{k}(S_{ij}) + u_{ij},
%\ese
We simulate discrete observations $W_{ij}$ from (\ref{eq:disc_obs}) where $U_{ij}$ are iid standard normal, and simulate $Y_{ij}$ from (\ref{tv_model}), where the measurement error $\epsilon_i(t)$ is generated from a mean zero Gaussian process with covariance $\cov\{\epsilon_i(s), \epsilon_i(t)\} = 2^{-|t-s|/5}$. For each subject, the observation time points $\{ S_{ij}\} $ and $\{T_{ij}\}$ are uniformly distributed on $\CT$ and independent from each other. We consider the following two settings for the time-varying coefficients:

\noindent \hskip20mm     \textit{Setting I}: $\beta_0(t) = 0.2t + 0.5$, $\beta_1(t) = \sin(\pi t/10)$; 

\noindent\hskip20mm     \textit{Setting \II}: $\beta_0(t) = t^{1/2}$, $\beta_1(t) = \sin(\pi t/5)$. 

\noindent We perform functional calibration using the \texttt{fdapace} package with $AIC$ as the principal component selection criterion. To fit a time-varying coefficients model after the functional calibration, we used the \texttt{tvLM} function in the R package \texttt{tvReg} which implements the kernel smoothing method in \cite{hoover1998nonparametric} and its built-in cross-validation procedure to choose the bandwidth. 
%By specifying a time grid on time domain $\mathcal{T} = [0,10]$, we can obtain $\beta_0(t) , \beta_1(t)$ for any $t \in \mathcal{T}$ for each simulation run. 
{As a comparison, we  consider the following estimators, i.e.,  the Oracle estimator with  the known synchronized true values of $X$, the KW estimator \citep{cao2015regression}, and the functional varying coefficients model (FVCM)  \citep{csenturk2013modeling}. The Oracle estimator is implemented by using the \texttt{tvReg} package,}
%Both the Oracle and LOCF estimators are calculated using the \texttt{tvReg} package, where LOCF simply uses the nearest last observation as surrogate of the true $X$ value. 
and the KW method for time-varying coefficient model is implemented by using the authors's own \texttt{asynchTD} function in the \texttt{AsynchLong} package. The FVCM method requires estimation of the covariance function of $X(t)$ and the cross-covariance function between $X(\cdot)$ and $Y(\cdot)$, which are calculated using the \texttt{fdapace} package. Bandwidths for all methods are selected using the built-in options of the packages mentioned above: generalized cross-validation of \texttt{fdapace}, cross-validation of \texttt{tvReg} and adaptive selection procedure of \texttt{AsynchLong}.

%\begin{table}[ht]
%    \centering
%    \caption{Comparison on procedures for different methods studied in simulations}
%    \begin{tabular}{c|c c}
%    \hline
%    \hline
%    Method & Description & Bandwidth Selection \\
%    \hline
%    FCART  & 1. functional calibration; & 1. GCV \\
%           & 2. estimate coefficients by (\ref{eqn:betat_est}). & 2. CV \\ 
%    LOCF  & 1. imputation by LOCF; & 1. NA \\
%           & 2. estimate coefficients by (\ref{eqn:betat_est}). & 2. CV \\ 
%    Oracle  & 1. use true values $X_i(T_{ij})$; & 1. NA \\
%           & 2. estimate coefficients by (\ref{eqn:betat_est}). & 2. CV \\ 
%    FVCM & 1. linear kernel smoother & 1. GCV \\
%          & 2. calculate coefficients & 2. NA \\
%    KW & kernel weighted method & adaptive selection \\
%    \hline
%    \end{tabular}
%    \label{tbl:sim:fcart:methods}
%\end{table}

We repeat the simulation 200 times for both settings and apply the proposed and competing methods to each data set. Following \cite{csenturk2010functional}, we compare different methods using two evaluation criteria: the mean absolute deviation error (MADE) and the weighted average squared error (WASE) 
\bse
\text{MADE}  =  \dfrac{1}{2|\mathcal{T}|} \sum_{r=0}^1 \dfrac{\int_{\mathcal{T}} |\wh \beta_r(t) - \beta_r(t)|dt}{\text{range}(\beta_r)}, \quad\quad
\text{WASE}  =  \dfrac{1}{2|\mathcal{T}|} \sum_{r=0}^1 \dfrac{\int_{\mathcal{T}} \{\wh \beta_r(t) - \beta_r(t)\}^2dt}{\text{range}^2(\beta_r)},
\ese
where $\text{range}(\beta_r)$ is the range of function $\beta_r(t)$, $r = 0, 1$.
%$|\mathcal{T}| = 10$. In practice, $\int_{\mathcal{T}}$ is the trapezoid rule numerical integration on $\mathcal{T}$. 
%As a integrated form on bias and squared error, MADE and WASE to are used to see how accurate and fluctuated time-varying estimates are in general.
%Details: each method estimate time-varying coefficient at the same time grid $\mathcal{T}$: 60 equally spaced points between minimum and maximum time points among $T_{ij}$ and $S_{ij}$.

\begin{figure}[htb]
	\centering
	\begin{tabular}{cc}
		\includegraphics[width = 2.6in, height=1.5in]{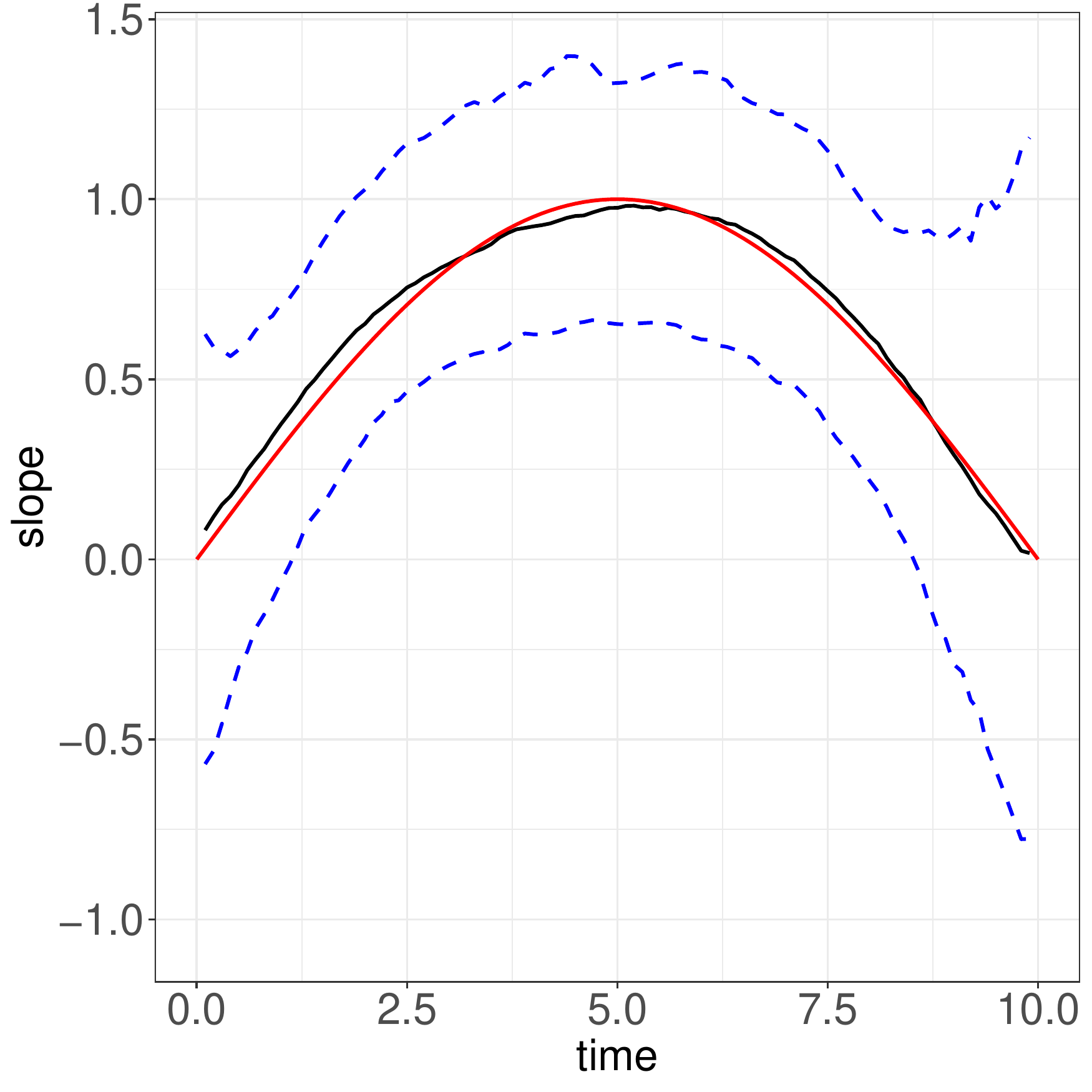}&
		\includegraphics[width = 2.6in, height=1.5in]{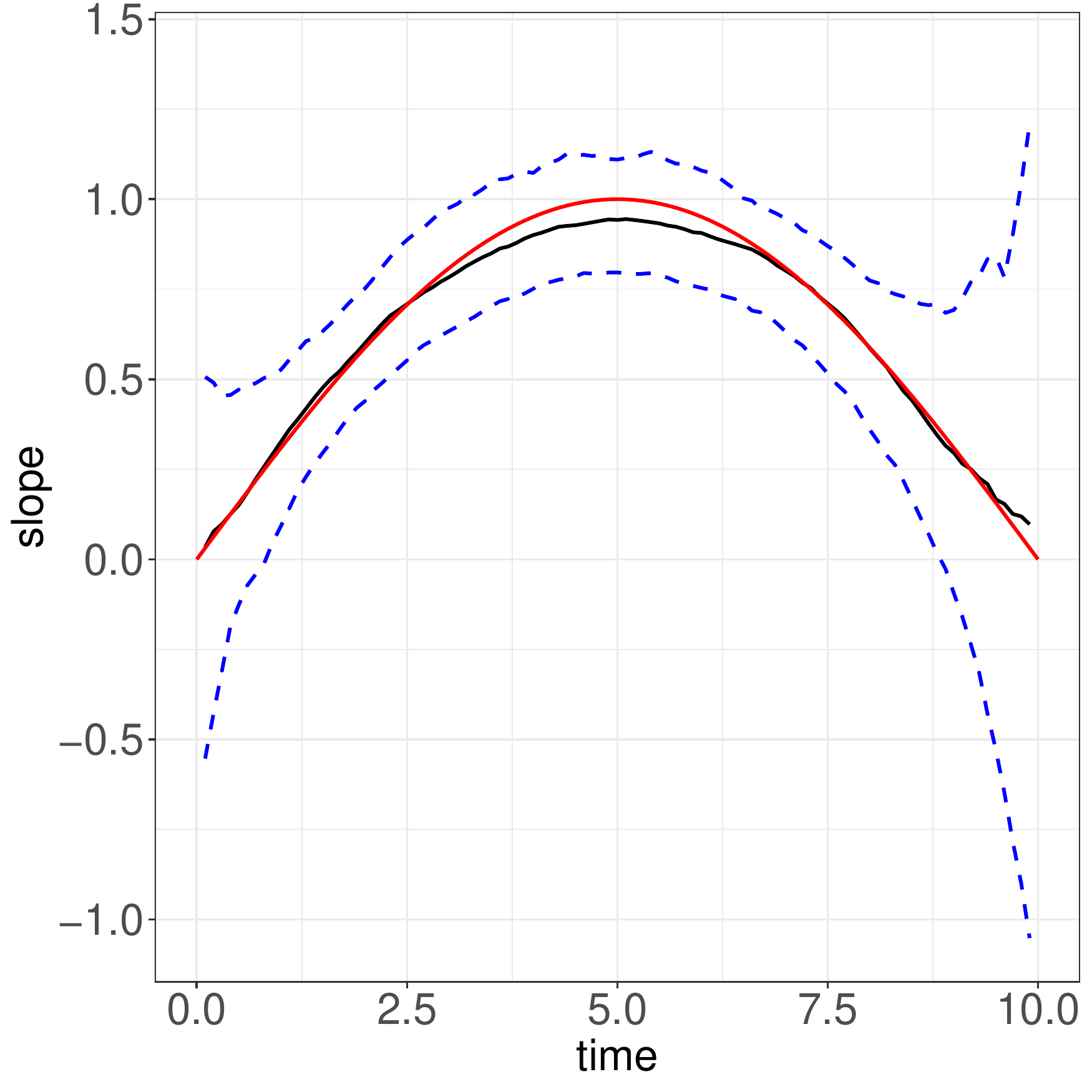}\\
				{FCAR} &{Oracle} \\
		 \includegraphics[width = 2.6in, height=1.5in]{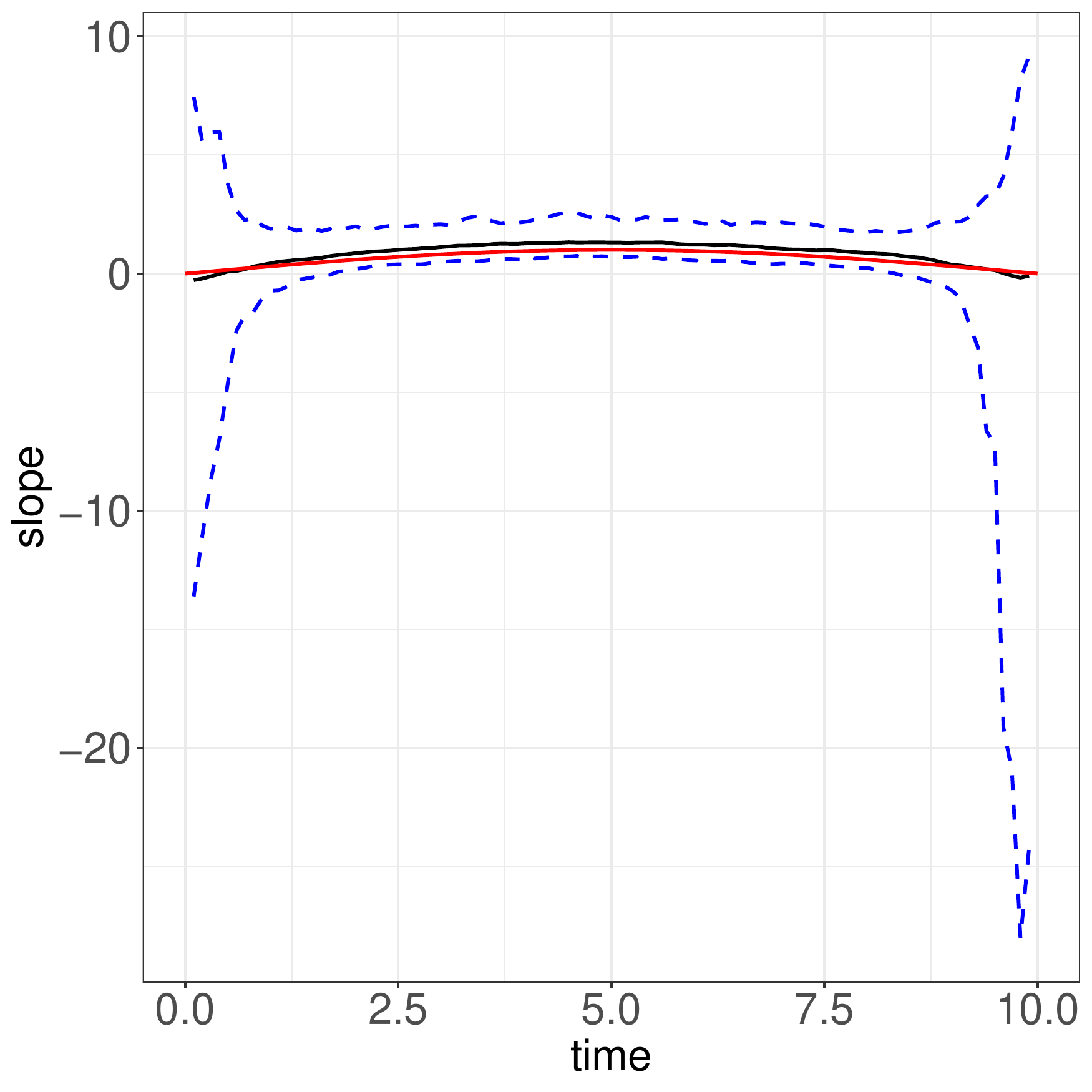} &
		\includegraphics[width = 2.6in, height=1.5in]{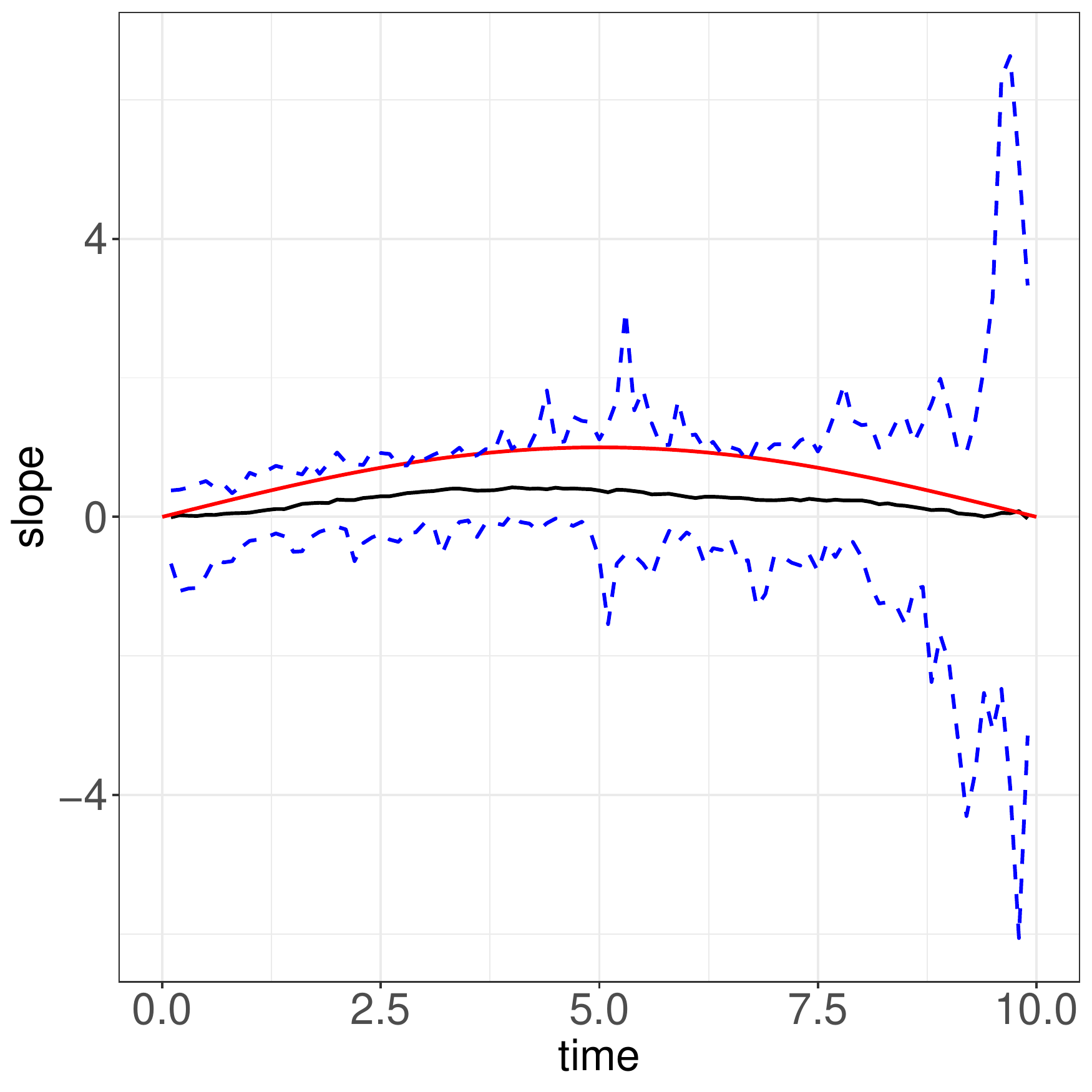}\\
		{FVCM} & {KW} \\
	\end{tabular}
	\vspace{1cm}
\caption{Summary of $\wh\beta_1(t)$ under Simulation 2, Setting I using various methods.  In each panel, black: median of $\wh \beta_1(t)$; red: true $\beta_1(t)$; dashed blue: 0.975 and 0.025 quantiles.} \label{fig:sim_fcart_setting1_b1}
\end{figure}

\begin{table}[ht]
	\centering
	\begin{tabular}{cccrccr}
		\hline 
		\hline 
		&Method & Criterion & Mean(SD) & Median & 25\% & 75\% \\
		\hline 
Setting I & FCAR & MADE & 0.319(0.151) & 0.302 & 0.207 & 0.388 \\ 
 & FVCM &  & 1.494(2.521) & 0.948 & 0.756 & 1.418 \\ 
 & KW &  & 1.452(3.533) & 1.026 & 0.868 & 1.297 \\ 
 & Oracle &  & 0.209(0.091) & 0.192 & 0.142 & 0.269 \\ 
\hline 
 & FCAR & WASE & 0.345(0.395) & 0.224 & 0.103 & 0.402 \\ 
 & FVCM &  & 461.495(5235.731) & 3.433 & 1.576 & 9.553 \\ 
 & KW &  & 1057.299(14234.414) & 3.830 & 1.716 & 15.221 \\ 
 & Oracle &  & 0.216(0.242) & 0.111 & 0.058 & 0.294 \\ 
    \hline 
		\hline 
Setting \II & FCAR & MADE & 0.263(0.104) & 0.246 & 0.186 & 0.321 \\ 
 & FVCM &  & 0.944(1.551) & 0.616 & 0.440 & 0.913 \\ 
 & KW &  & 1.153(1.669) & 0.720 & 0.605 & 1.115 \\ 
 & Oracle &  & 0.180(0.062) & 0.172 & 0.137 & 0.220 \\ 
\hline 
 & FCAR & WASE & 0.316(0.351) & 0.200 & 0.093 & 0.394 \\ 
 & FVCM &  & 299.501(3746.442) & 1.464 & 0.613 & 5.921 \\ 
 & KW &  & 204.128(1538.798) & 2.709 & 1.196 & 15.833 \\ 
 & Oracle &  & 0.287(0.347) & 0.183 & 0.088 & 0.330 \\ 
    \hline 
		\hline 
	\end{tabular}
	\vspace{1cm}
	\caption{Simulation 2: MADE and WASE of various methods.  SD: standard deviation);  25\%:   25\% quantile; 75\%: 75\% quantile.}
	\label{tbl:sim:fcart:compare}
\end{table}
As summarized in Table~\ref{tbl:sim:fcart:compare},  the proposed FCAR method yields MADE and WASE that are close to the Oracle estimator. The FVCM and KW methods equipped with  built-in tuning parameter selectors perform  worse than FCAR, {likely because both of them, by evoking bivariate kernel smoothing while estimating univariate coefficient functions in Model (\ref{tv_model}),  are numerically unstable.} {Both FVCM and KW present large mean WASEs
%because of the boundary effects  
(Fig~\ref{fig:sim_fcart_setting1_b1}), which are further magnified  by the square operator.} %It is further enlarged by an outlier in setting I as shown in Fig~\ref{fig:sim_fcart_setting1_b1}}

%\yhlcomment{FVCM and KW are pretty bad even after removing the boundary points. Need to think about this sentence.}

MADE and WASE are overall numerical summaries combining $\wh \beta_0(\cdot)$ and $\wh \beta_1(\cdot)$; we also provide graphical summaries of $\wh \beta_0(\cdot)$ and $\wh \beta_1(\cdot)$ separately. \yhl{In Figure \ref{fig:sim_fcart_setting1_b1}, we summarize $\wh \beta_1(t)$ under Setting I by all 4 methods mentioned above; a similar graph (Figure C.1) under Setting \II ~is provided in the online Appendix.}
%As seen, the LOCF estimator for $\beta_1(t)$ is significantly biased toward 0, due to the well-known attenuation effect in the measurement error literature \citep{Carroll2006}. 
The proposed FCAR estimator for $\beta_1(t)$ has negligible biases and overall performance comparable to the Oracle estimator. In contrast, with slower convergence rates and numerical instability of bivariate kernel smoothing, the KW and FVCM estimators for $\beta_1(t)$ are highly variable and affected by the boundary effect.
Graphical summaries of $\wh \beta_0(t)$ allude to the same message. We therefore relegate the graphs on $\wh \beta_0(t)$ under these two settings to Figures C.2 and C.3 in Appendix C. \yhl{Across both settings, FCAR takes an average of 0.45 minutes to estimate the time-varying coefficient functions. A typical run of the bootstrap procedure takes an additional 8.47 minutes under a 10-core parallel. In contrast, KW takes a combined 5.18 minutes for estimation and inference. The FVCM takes an average of 1.93 minutes in estimation, and relies on a bootstrap procedure similar to ours to make inference.}

\yl{As KW and FVCM are visually sensitive to the boundary effects, we furnish  summary tables and plots with the 95\% truncated domains in Appendix C for a more fair comparison.} \yl{In addition, we consider the case of MEF with an added sparsity level of $m_i = 15$, which again demonstrates the fine performance of the proposed
FCAR estimator; see  Appendix C}.

%\begin{table}[ht]
%	\centering
%	\caption{MADE and WASE statistics: mean value (sample standard deviation), median, 25\% quantile, 75\% quantile in 200 simulations under setting I.}
%	\begin{tabular}{cccccc}
%		\hline 
%		\hline 
%		Method & Metric & Mean & Median & 25\% & 75\% \\
%		\hline 
%		FCART & MADE & 0.3188(0.1511) & 0.3017 & 0.207 & 0.3885 \\ 
%		FVCM &  & 1.4937(2.5206) & 0.9483 & 0.7556 & 1.4176 \\ 
%		KW &  & 1.4524(3.5327) & 1.026 & 0.8684 & 1.297 \\ 
%		LOCF & & 0.9117(0.0483) & 0.9147 & 0.8755 & 0.943 \\ 
%		Oracle & & 0.2088(0.091) & 0.1923 & 0.1424 & 0.2693 \\
%		\hline 
%		FCART & WASE & 0.3455(0.3953) & 0.2244 & 0.1028 & 0.4022 \\ 
%		FVCM & & 461.4947(5235.7312) & 3.4327 & 1.5763 & 9.553 \\ 
%		KW &  & 1057.2985(14234.4135) & 3.8304 & 1.7165 & 15.221 \\ 
%		LOCF & & 1.3555(0.1367) & 1.3511 & 1.2692 & 1.4499 \\ 
%		Oracle & & 0.2155(0.2425) & 0.111 & 0.0575 & 0.2937 \\
%		\hline 
%	\end{tabular}
%	\label{tbl:sim:fcart:compare1}
%\end{table}
%As summarized in Table~\ref{tbl:sim:fcart:compare1}, we can find FCART has much lower MADE and WASE and smaller variation than FVCM and KW methods under setting I. The extremely high WASE of KW and FVCM is an indicator of unstable computation procedures and varied estimates.

% \begin{figure}[htb]
% 	\centering
% 	\begin{tabular}{cc}
% 		\includegraphics[width = 2.6in, height=1.5in]{./simulation/FCAR_beta1_sim_xt2_bands_100.pdf}&
% 		\includegraphics[width = 2.6in, height=1.5in]{./simulation/Oracle_beta1_sim_xt2_bands_100.pdf}\\
% 				{FCAR} &{Oracle} \\
% 		\includegraphics[width = 2.6in, height=1.5in]{./simulation/FVCM_beta1_sim_xt2_bands_100.pdf} &
% 		\includegraphics[width = 2.6in, height=1.5in]{./simulation/KW_beta1_sim_xt2_bands_100.pdf}\\
% 		{FVCM} & {KW}\\
% 	\end{tabular}
% 	\vspace{1cm}
% \caption{Summary of $\wh\beta_1(t)$ under Simulation 2, Setting II using various methods.  In each panel, black: median of $\wh \beta_1(t)$; red: true $\beta_1(t)$; dashed blue: 0.975 and 0.025 quantiles.} \label{fig:sim:fcart:all2:b1}
% \end{figure}

%
%
%
%
%
%
%
%
%
%
%
%
%
%
%

%
 
\section{Real Data Analysis} \label{sec:realdata}

We  apply the proposed FCAR method to 
%the Study of Women's Health Across the Nation (SWAN)  
the SWAN data described in Section \ref{sec:introduction}. 
%which can be downloaded from \url{https://www.icpsr.umich.edu/web/ICPSR/series/00253}. 
%SWAN \citep{bromberger2010longitudinal} was a large longitudinal observational study on women's health during menopausal transition, which plays an essential role in elderly women's lives and is associated with changes %in physiological performance such as
%in hormone levels, cardiovascular risk factors and other physical measures. 
%with the goal to monitor the physical, biological, psychological, and social changes for women during their middle years. The menopausal transition plays an essential part in elderly women's lives and is associated with changes in physiological performance such as hormone levels, cardiovascular risk factors and physical measures. 
The study admitted 3,302 premenopausal or early perimenopausal women between 1996 and 1997, with the baseline age ranging from 42 to 53. 
%through seven designated research centers.  participants represent five racial/ethnic groups and aged  in the initial wave (baseline visit). 
These women were scheduled to have annual followups up to 10 years, although various hormonal, physical and cardiovascular biomarkers were measured according different schedules as illustrated in Figure \ref{fig:days_id} until the study ended in 2008. 
%See \cite{bromberger2010longitudinal} for more details of the study.
%to evaluate the association between follicle-stimulating hormone and triglycerides adjusting for the body mass index. We also compare analysis results with the last observation carried forward (LOCF). It is demonstrated that FCART has a favorable performance and leads to convincible results in regression analysis of asynchronous longitudinal data. 
%\subsection{Follicle-stimulating Hormone in the Menopausal Transition}
% introduction of menopause and FSH
%Menopausal status was based on menstrual bleeding patterns in the previous 12 months and was categorized as premenopausal, early perimenopausal, late perimenopausal and postmenopausal \citep{bromberger2010longitudinal}.
One of the most important biomarkers in menopausal studies is the follicle-stimulating hormone (FSH) level,  the  outcome variable of our primary interest. 

As declining follicular reserve is the immediate cause of the perimenopausal and menopausal transitions \citep{richardson1987follicular}, an increase in the serum FSH level was  one of  the major endocrine changes associated with  menopausal transitions \citep{burger1995endocrinology}. FSH levels rise progressively before the final menses and will continue for 2--4 years, before remaining elevated postmenopause \citep{burger1999prospectively}.
%Change in FSH levels across the menopausal transition has been an active area in medical research: 
Changes in FSH levels have been linked to or are precursors of various medical conditions. For example, abnormal variations of FSH levels are related to the depressive symptoms during the menopausal transition \citep{bromberger2010longitudinal}, and may also increase women's risk of developing cardiovascular disease after menopause \citep{el2016trajectories}.

%Some recent work includes but not limited to the following. Both \cite{randolph2011change} and \cite{randolph2004change} analyzed the SWAN dataset, and reached similar conclusions: change in serum FSH across the menopausal transition were related to race/ethnicity. 
%Similarly,  \cite{tepper2012trajectory} ascertains that race/ethnicity and body mass index affect the trajectory of FSH change during the menopausal transition. 
%\cite{park2017association} investigates the association between changes in oestradiol and follicle-stimulating hormone levels, and \cite{bromberger2010longitudinal} focuses on the connection between serum hormone levels and depressive symptoms across the menopausal transition. \cite{el2016trajectories} concludes that the change could potentially increase women's risk of developing cardiovascular disease after menopause.
%Then \cite{serviente2019follicle} and \cite{wang2020follicle} study the relationship between FSH and lipid metabolism in postmenopausal women and perimenopausal women, respectively.

% what we will 
Therefore, studying the dynamic relationship between FSH and other physiological measurements is of great importance to understand women's reproductive life and their midlife health \citep{bromberger2010longitudinal}. Following \cite{wang2020follicle}, we study the association between FSH and triglycerides (TG) adjusting for age, income level and body mass index (BMI). FSH was measured every year for the SWAN participants following the hormone measurement schedule in Figure \ref{fig:days_id}, whereas TG and BMI were following the cardiovascular and physical measurement schedules in Figure \ref{fig:days_id}. Of note, TG was not collected in year 2 or beyond year 8, and 47.5\% of BMI measurements were asynchronous with FSH. We also included the baseline age and income as time-invariant covariates, where the income was dichotomized (1 if annual income is more than \$50k and 0 otherwise). After removing subjects with missing incomes, there are 1,634 high income subjects and 1,578 low income subjects in the data. We focus our analysis on the first 8 years of the study, when FSH, TG and BMI are all available. An added rationale behind this truncation is that all participants experienced the entire menopausal transition by year 8,  becoming postmenopausal or late perimenopausal afterwards \citep{bromberger2010longitudinal}. 

Existing works, such as \cite{wang2020follicle}, assume the association between FSH and the covariates are time-invariant and ignore asynchronous issue in this data set, {which may mask some intriguing time-varying associations. }
%Considering FSH as a time-varying variable, FCAR can be applied to the SWAN data to obtain coefficients as a function in time while taking asynchronous observations into account.  
Instead, we apply the time-varying coefficients model  to model the dynamic relationship between FSH and other time varying or invariant covariates.
Among the competing methods described in Section \ref{sec:simulation:fcart}, the kernel weighted estimator (KW) requires that all time-varying covariates are measured at the same time points, which is not applicable in our data since TG and BMI are measured on different time as well; the functional varying coefficient model (FVCM) of \cite{csenturk2010functional} was proposed for univariate time-varying covariates and is not readily applicable to multiple time-varying covariates in this data set. 

{To accommodate two time-varying covariates in our data, we slightly extend the proposed FPCA  to a multivariate setting as described in Appendix B and implement it} by using the \texttt{fdapace} package in R, where a built-in generalized cross-validation (GCV) procedure is used to select the bandwidths for mean and cross-covariance estimations. We then use the conditional AIC described in (\ref{eq:aic}) to select the number of principal components for TG and BMI separately. To implement the undersmoothing scheme described in condition \ref{cond:tv_bandwidth}, we multiply the GCV selected bandwidths by a factor of $n^{-1/10}$, refit FPCA using the undersmoothing bandwidths, and use the FPCA calibrated values {for the subsequent analyses}. 

\begin{figure}[htb]
	\begin{subfigure}{0.48\textwidth}
		\centering
		\includegraphics[width = 0.8\textwidth]{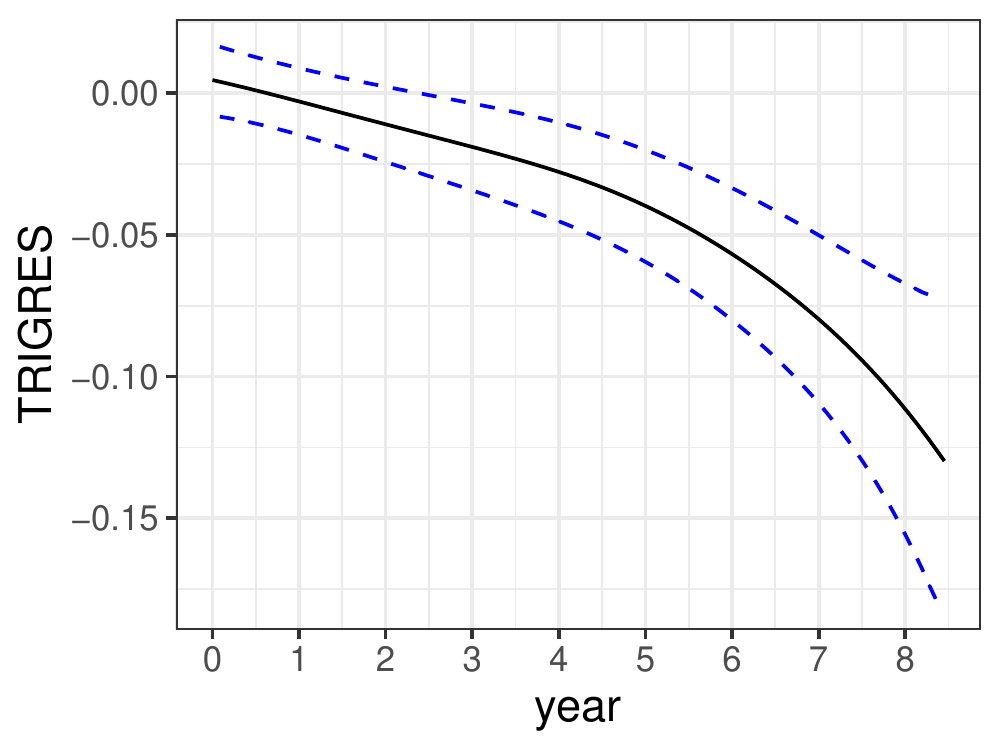}
		\caption{Triglycerides}
		\label{fig:realdata:tg:fcar}
	\end{subfigure}
% 	\begin{subfigure}{0.48\textwidth}
% 		\centering
% 		\includegraphics[width = 0.8\textwidth]{real_data/twoxt_z_LOCF_TRIGRES_slope_bts_income_age.pdf}
% 		\caption{LOCF}
% 		\label{fig:realdata:tg:locf}
% 	\end{subfigure}
	\begin{subfigure}{0.48\textwidth}
		\centering
		\includegraphics[width = 0.8\textwidth]{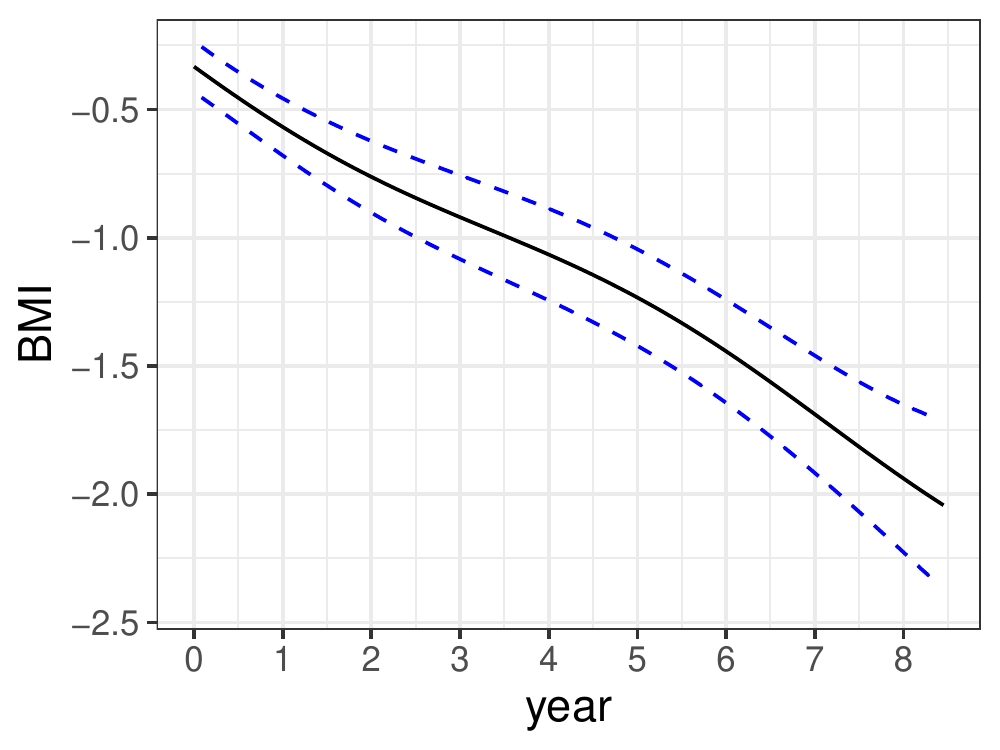}
		\caption{BMI}
		\label{fig:realdata:bmi:fcar}
	\end{subfigure}
% 	\begin{subfigure}{0.48\textwidth}
% 		\centering
% 		\includegraphics[width = 0.8\textwidth]{real_data/twoxt_z_LOCF_BMI_slope_bts_income_age.pdf}
% 		\caption{LOCF}
% 		\label{fig:realdata:bmi:locf}
% 	\end{subfigure}
	\begin{subfigure}{0.48\textwidth}
		\centering
		\includegraphics[width = 0.8\textwidth]{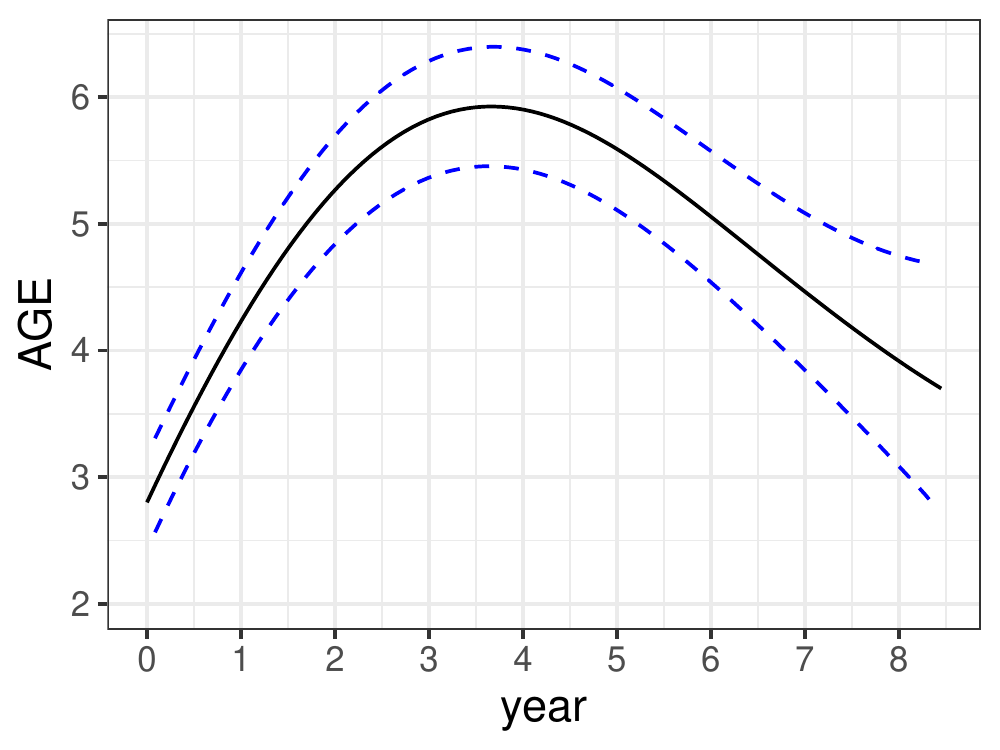}
		\caption{Baseline Age}
	\end{subfigure}
	\begin{subfigure}{0.48\textwidth}
		\centering
		\includegraphics[width = 0.8\textwidth]{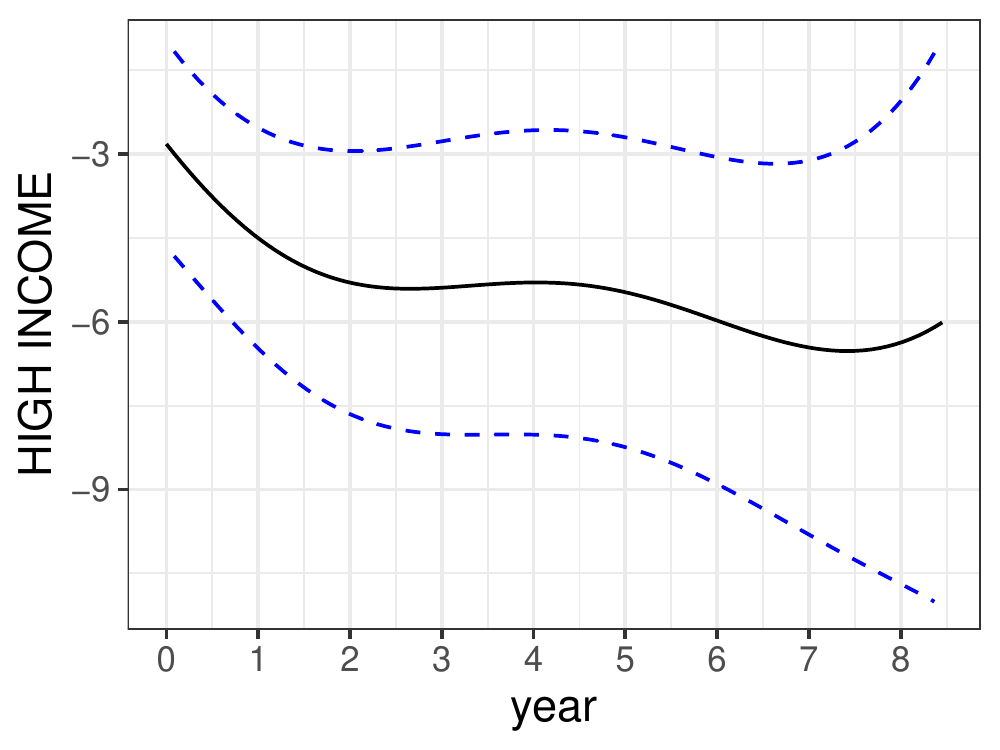}
		\caption{Income}
	\end{subfigure}
	\caption{SWAN data analysis using FCAR: time-varying coefficient model on FSH against TG, BMI, age and income. %Top row: coefficient function for TG; bottom row: coefficient function for BMI. Left column: FCAR; right column: LOCF. 
	In each panel, the solid curve is the estimated coefficient function and the dashed curves are 95\% point-wise confidence intervals obtained using bootstrap.}
	\label{fig:realdata}
\end{figure}

%To follow the undersmoothing condition, we choose bandwidths as $n^{-1/10}$ multiplying the optimal chosen by generalized cross-validation in \texttt{fdapace} package. Then bandwidths in functional calibration are set as $h_{\mu} = 0.76, h_R = 1.33$ for triglycerides (TG) and $h_{\mu} = 1.28, h_R = 1.15$ for BMI respectively. 
%Both the LOCF and FCAR methods use the local polynomial method of \citep{hoover1998nonparametric} to estimate the time-varying coefficients in Model (\ref{eq:mv_tv_model}). To make the results comparable, we use the same bandwidth to fit Model (\ref{eq:mv_tv_model}). %\xc{which is the determined by cross-validation using LOCF}. 
%where bandwidth $h = 1.63$, . Both functional calibration and time-varying coefficient estimation use the Gaussian kernel. We use $AIC_{cond}$ to select the number of principal components, which performs better than marginal likelihood $AIC$ for real data with large measurement error. 
\yhl{We  regress FSH against the calibrated TG and BMI values and adjust for time-invariant covariates age and income, using a multivariate time-varying coefficient model
\bse
    Y_i(t) =\beta_0(t) +\bdbeta_z\trans(t)  \BZ_i + \bdbeta_x\trans(t) \BX_i(t)+\epsilon_i(t).
\ese
Fig \ref{fig:realdata} shows the estimated coefficient functions for TG, BMI, age and income using FCAR, respectively, where the 95\% pointwise confidence intervals are obtained using bootstrap.} As commented in Remarks 2 and 3, we resample the subjects, perform FPCA using the same bandwidth as in real data to every bootstrap sample in order to properly take into account the FPCA estimation errors. The pointwise confidence intervals in Fig \ref{fig:realdata} are based on a normal approximation suggested by Theorem \ref{thm:beta_t}, where the pointwise standard error is estimated based on 200 bootstrap replicates. 
%For convenience and faster computation, we fix the bandwidths at the choice from the original model fitting. 

%
%
%
%
%
%
%
%
%
%
%
%
%
%
%
\yhl{The estimated time varying coefficients reveal that FSH is negatively associated with TG, which is consistent with the SWAN data analysis conducted in \cite{el2016trajectories} using time-invariant regression models. %\cite{wang2017follicle} and 
\cite{wang2020follicle} studied the association between FSH and TG among postmenopausal and perimenopausal women separately using independent studies. The comparison between their results suggests a stronger negative association between FSH and TG among postmenopausal women than perimenopausal women, which supports our findings in Figure \ref{fig:realdata} (a) that the negative association between FSH and TG becomes stronger through the menopausal transition.
Similarly, Figure \ref{fig:realdata}(b) suggests that FSH is negatively associated with BMI. This is consistent with previous findings in the SWAN literature 
%\cite{serviente2019follicle} and 
\citep{randolph2004change}, which  
%classified the longitudinal observations by menopausal status of the women (premenopausal, early perimenopausal, late premenopausal and postmenopausal), and their cell mean models 
suggested that the negative association between FSH and BMI becomes stronger throughout the menopausal transition. This time-varying effect of BMI on FSH is not only confirmed by our study, but can be visualized as a continuous curve in Figure  \ref{fig:realdata}(b).}

\yhl{
Time-invariant variables, such as age and income, are confounders, whose effects  need to be adjusted for in the model. The effect of baseline age represents a cohort effect, reflecting different baseline FSH levels in different age cohorts. 
The positive association between age and FSH  seen in Figure \ref{fig:realdata}(c) is  consistent with the literature that FSH is elevated as women age through the menopause transition \citep{burger1999prospectively}.
%, tepper2012trajectory} 
%and primary ovarian insufficiency \citep{DeVos2010}.
Also we find that lower income women are more likely to present higher FSH, agreeing to the literature that links low socioeconomic status to high FSH \citep{wise2002lifetime},
%silven2022incidence}, 
possibly because of poor health awareness
 \citep{burger1995endocrinology}, risk behaviors \citep{haddad2008hpv16},  and inadequate access to health care \citep{barut2016there}.}

\section{Conclusions}\label{sec:conclusion}
We have proposed a new functional calibration method, termed Functional Calibration for Asynchronous Regression (FCAR), for learning sparse asynchronous longitudinal data. The key idea behind the approach is to calibrate the missing synchronized covariates by the functional principal component analysis (FPCA) approach, which can be easily implemented using existing software. More broadly, our method is applicable to asynchronous longitudinal regression with time-invariant or time-varying coefficients, and addresses a serious limitation of the existing literature. 
%that the naive LOCF method leads to biased estimator of the coefficients, which are attenuated toward 0. 
Indeed, our FCAR estimator in a time-invariant regression model enjoys nice asymptotic properties, such as root-$n$ consistency and asymptotic normality.
%; our estimator for a time-varying coefficient model enjoys the same optimal convergence rate as those in the literature, such as \cite{Cai2000jasa}. 
By implementing an undersmoothing scheme in our functional calibration method, the FPCA estimation errors cause a negligible bias to the estimated model, but will inflate the asymptotic variance of the final estimator. Based on these theoretical findings, we recommend to  use bootstrap standard error that takes into account FPCA errors, rather than using naive standard errors. Our theoretical analysis as well as our empirical studies show that our proposed method outperforms the existing methods, including the kernel weighted estimator of \cite{cao2015regression} and the FVCM method of \cite{csenturk2010functional}.

%We start by introducing asynchronous longitudinal data, pointing out the difficulties and drawbacks of existing methods. To improve slow converge rates and computation stability, we propose to separate the procedure into two steps: using FPCA to reconstruct random processes and do imputation, then fitting a regression model to functional calibrated data. Because the imputation error is actually reduced again in the second step of model fitting, this is more efficient than using one step, which usually involves the bivariate kernel. As shown in theory and simulation studies, the proposed method provides a more accurate and stable estimator than other available methods for estimating either time-invariant coefficients or time-varying coefficients.

\yhl{As a reviewer pointed out, the methods of \cite{cao2015regression} and \cite{csenturk2013modeling} can handle generalized outcomes, whereas our investigation has been confined to Gaussian type responses. This extension requires substantial theoretical  work and we defer it to future work.} As demonstrated in Appendix B and our real data analysis, 
the proposed FCAR method can be easily implemented for multiple asynchronous time-varying covariates, when the number of time-varying covariates is not too high. 
When the number is high, the computational load of multivariate FPCA (mFPCA) can rapidly escalate, leading to an unmanageable number of cross-covariance functions to estimate and causing the mFPCA estimators to become unreliable. In these situations, choosing the pertinent time-varying covariates presents a formidable challenge of `model selection with error-in-variable,' as the calibrated covariate values are subject to estimation errors, which are not independent and possess complex structures. These difficulties deserve further exploration.
%When this number is high, the computation burden of multivariate FPCA (mFPCA) can quickly grow out of control as we will have too many cross-covariance functions to estimate, and the mFPCA estimators may become unstable. In this setting, selecting the relevant time-varying covariates becomes a challenging `model selection with error-in-variable' problem, because the calibrated covariate values contain estimation errors that are not independent and have complicated structures. These challenges also warrant future investigations.

%  The \backmatter command formats the subsequent headings so that they
%  are in the journal style.  Please keep this command in your document
%  in this position, right after the final section of the main part of 
%  the paper and right before the Acknowledgements, Supporting Information (Supplementary %  Materials),   and References sections. 

\backmatter

%  This section is optional.  Here is where you will want to cite
%  grants, people who helped with the paper, etc.  But keep it short!

\section*{Acknowledgements}
We thank the Editor, the anonymous AE and referee for their insightful comments and suggestions that have improved substantially  the quality of the manuscript. The work is partially supported by grants from the National Institutes of Health.
The data and code are  available at \url{https://github.com/chxyself25/Functional\_Calibration}.  
\section*{Supporting Information}

Web Appendices A--C, referenced in the main text, are available with
this paper at the Biometrics website on Wiley Online
Library.\vspace*{-8pt}

%As with most nonparametric methods, the functional calibration approach includes several bandwidth selection procedures. Compared to the bandwidth used in FPCA, We found that the result is more sensitive to the bandwidth used in time-varying coefficient estimation. The proposed method needs three procedures for selecting optimal bandwidth for estimating the time-varying coefficient. The computation time could be one concern in practice. In addition, regression with one covariate is the simplest case for longitudinal data studies. Although our approach can be easily generalized to multiple time-varying covariate models, there are concerns regarding both two steps. For functional calibration, multivariate functional principal component analysis needs to be adopted and will cause more computation difficulties. For the later step of fitting regression models, regularization needs to be applied for variable selection and estimation. 

%\bibliographystyle{acm} % use for numbered citations along with options given in preamble. Look at the main thesis.tex file
\bibliographystyle{biom} 
\bibliography{async1_ref.bib}

%  Here, we create the bibliographic entries manually, following the
%  journal style.  If you use this method or use natbib, PLEASE PAY
%  CAREFUL ATTENTION TO THE BIBLIOGRAPHIC STYLE IN A RECENT ISSUE OF
%  THE JOURNAL AND FOLLOW IT!  Failure to follow stylistic conventions
%  just lengthens the time spend copyediting your paper and hence its
%  position in the publication queue should it be accepted.

%  We greatly prefer that you incorporate the references for your
%  article into the body of the article as we have done here 
%  (you can use natbib or not as you choose) than use BiBTeX,
%  so that your article is self-contained in one file.
%  If you do use BiBTeX, please use the .bst file that comes with 
%  the distribution.  In this case, replace the thebibliography
%  environment below by 
%
%  \bibliographystyle{biom} 
% \bibliography{mybibilo.bib}

%  If your paper refers to supporting web material, then you MUST
%  include this section!!  See Instructions for Authors at the journal
%  website http://www.biometrics.tibs.org

%\appendix

%  To get the journal style of heading for an appendix, mimic the following.

%\section{}
%\subsection{Title of appendix}

%Put your short appendix here.  Remember, longer appendices are
%possible when presented as Supplementary Web Material.  Please 
%review and follow the journal policy for this material, available
%under Instructions for Authors at \texttt{http://www.biometrics.tibs.org}.

\label{lastpage}

\end{document}